\documentstyle[prl,aps,twocolumn,epsf]{revtex}
\begin{document}
\draft
\title{Theory of
Phonon-Assisted Multimagnon Optical Absorption and Bimagnon States
in Quantum Antiferromagnets.}
\author{J. Lorenzana\cite{pra} and G. A. Sawatzky}
\address{
Laboratory of Applied and Solid State Physics, Materials Science Centre,\\
University of Groningen, Nijenborgh 4, 9747 AG Groningen, The Netherlands}
\date{\today}
\maketitle
\begin{abstract}
We calculate the effective charge for multimagnon  infrared (IR)
absorption assisted by phonons in a perovskite like antiferromagnet and
we compute the spectra for two magnon absorption using
interacting spin-wave theory. The full set of equations for the
interacting two magnon problem is presented
 in the random phase approximation  for arbitrary total momentum
of the magnon pair.  The spin wave theory results fit very
well the primary peak of recent measured bands in the
 parent insulating compounds of cuprate superconductors.
The line shape is explained as being due to the absorption of
one phonon  plus a new quasiparticle excitation  of the
Heisenberg Hamiltonian that consists off a long lived virtual bound state of
 two magnons (bimagnon).
The bimagnon states have  well defined energy and momentum
in a substantial portion of the Brillouin zone.
The higher energy bands are explained as
one phonon plus higher multimagnon absorption processes.
Other possible experiments for observing bimagnons are proposed.
In addition we predict the line shape for the spin one system
La$_2$NiO$_4$.
\end{abstract}
\narrowtext
\section{INTRODUCTION}
In the late 50's Newman and Chrenko presented
a pioneering infrared absorption study in NiO\cite{new59}.
Among other results they showed a band at 0.24eV that  correlated with
the disappearance of antiferromagnetism above the N\'eel temperature
and hence of likely magnetic origin.
This was very  puzzling because in the NaCl
structure, with an O at an inversion center,   a
 direct magnetic absorption is not allowed\cite{tan65}.
In fact in a typical two magnon excitation,
like for example, two spin flips on adjacent metals atoms,
 all relaxation of charge in the excited state is symmetric with no net
 dipole moment.
An explanation of this puzzle was given by Mizuno and Koide\cite{miz63}
who observed that the joint absorption process of a phonon
and two magnons is allowed since then the symmetry of the lattice, after
the phonon excitation, is effectively lower. To the best of our knowledge no
detailed theory existed of this effect. Note that the Mizuno and
Koide\cite{miz63} paper is from the early days of  Anderson's superexchange
theory.

Also very puzzling data was recently presented by
Perkins {\it et al}\cite{per93,per94}.
They measured the absorption in many different parent insulating
compounds of high-T$_{\rm c}$ superconductors.
The data shows a narrow primary peak in the charge transfer gap  and a
set of side bands.
Initially the narrow peak was associated with an unidentified exciton.
However no exciton is expected in this energy range\cite{esk90,mcm90}.

The propose of this paper is to present a detailed
theory of phonon-assisted multimagnon infrared absorption.
We apply the theory to layered insulators and
show that it explains  Perkins {\it et al.}
data (Sec.~\ref{com}). We estimate the coupling constant of light
with multimagnon excitations (Sec.~\ref{eff})
and we calculate the line shape for one phonon plus two magnon  absorption
using interacting spin wave theory.
For this we need to solve the two-magnon problem for
arbitrary total momentum. We present the full set of equations
in the random-phase-approximation (RPA)
 and solve them approximately  (Sec.~\ref{lin}).
 The narrow primary peak reproduced
in Fig.~\ref{ade} is explained in terms
of a new quasiparticle excitation. It consists
of a long lived virtual bound state of magnons referred to here as a
bimagnon. The new state has  well defined energy and momentum
in a substantial portion of the Brillouin zone.
We also present a prediction for the line shape of the spin one system
La$_2$NiO$_4$.

So far complementary information on  antiferromagnetism
in perovskite materials
 has come from probes like  neutron scattering and Raman
light scattering\cite{man91}. Our results show that
IR absorption can be used to study magnetic properties of this systems.
 This is interesting because the technique is relatively simple.

In the conclusions (Sec.~\ref{conc}) we discuss and
 propose other experiments which
should be able to detect the   new excitations.

A short account of this results was presented elsewhere\cite{lor95}.

\section{THEORY OF PHONON-ASSISTED IR ABSORPTION OF MAGNONS}
\label{eff}

\subsection{Model Hamiltonian}

We concentrate on the case of
a spin $S=1/2$ and  2 dimensional (2-d) material like the
Cu-O layers of the cuprates
 but we indicate the corresponding generalizations
for a system with larger dimension and/or spin like La$_2$NiO$_4$
or NiO.  We consider
a  three-band-Peierls-Hubbard model\cite{yon92,yon93}
in the presence of an electric field ($\bbox{ E}$),
  \begin{eqnarray}
  H&=&  \sum_{i\neq j, \sigma} t_{ij}(\{ \bbox{u}_k\}) c_{i \sigma}^\dagger
  c_{j \sigma}
   + \sum_{i,\sigma} e_{i}(\{ \bbox{u}_k\},\bbox{ E})  c_{i \sigma}^\dagger
      c_{i \sigma} \nonumber \\
   & &+ \sum_{i} U_{i} c_{i \uparrow}^\dagger  c_{i \downarrow}^\dagger
		       c_{i \downarrow}        c_{i \uparrow}
   + \sum_{\langle i\neq j\rangle, \sigma,\sigma'} U_{ij} c_{i\sigma}^\dagger
	   c_{j \sigma'}^\dagger c_{j \sigma'} c_{i \sigma}
\nonumber \\
   & &+\sum_l \frac{1}{2M_l}\bbox{P}_l^2
      +\sum_{k,l} \frac12 \bbox{u}_k {\cal K}_{kl} \bbox{u}_l
-\bbox{ E} \bbox{ P}_{\rm ph}.
\label{ham}
\end{eqnarray}
Here, $c_{i \sigma}^\dagger$ creates a {\em hole\/} with spin $\sigma$
at site $i$ in the Cu $d_{x^2-y^2}$ or the O $p_{x,y}$ orbital.
For simplicity Cu atoms are kept fixed and O atoms
 are allowed to move with displacements $\bbox{u}_k$.
These restrictions will
be effectively removed when we compare with real experiments
since measured phonon properties will be used.
${\cal K}_{kl}$ is a spring constant tensor and $\bbox{P}_l$
is the  momentum canonically conjugate to  $\bbox{u}_l$,
$\bbox{ P}_{\rm ph} = Ze\sum_{i}\bbox{u}_k$ is the phonon dipolar moment
and $Z$ is the ionic charge of O ($Z=-2$).
 For electron-lattice coupling, we assume that the nearest-neighbor
Cu-O hopping is modified by the O-ion displacement $ \bbox{u}_k$ as
$t_{ij}=t\pm\alpha  \bbox{\delta u}_k$, where $\bbox{\delta}$
is a unit vector in the direction of the corresponding Cu-Cu
bond, $\bbox{\delta}=\bbox{\hat{x}}$, $\bbox{\hat{y}}$
(and $\bbox{\hat{z}}$ for a 3-d system) and  the $+$ ($-$)
applies if the Cu-O bond shrinks (stretches) with positive $\bbox{\delta u}_k$
(O-O hopping is neglected here). $t$ is given by $1/2\sqrt{3}pd\sigma$
in terms of Slater-Koster integrals\cite{sla54}.
 The site energy contains the coupling to the electric
field and in addition the Cu site energy  is
  assumed to be modulated linearly by the displacements of
the O ions,
\begin{eqnarray}
\label{eicu}
 &e_i&= E_d + \beta \sum_{k} (\pm) \bbox{\delta u}_k
+e \bbox{ E r}_i\ \ \ ({\rm Cu}),\\
&e_i&= E_p + e \bbox{ E r}_i\ \ \ ({\rm O}) .
\end{eqnarray}
In Eq.~(\ref{eicu})  the sum extends over
 the four surrounding O ions. The sign takes the value $+$ ($-$)
if the bond becomes longer (shorter) with positive $\bbox{\delta u}_k$
 and $ \bbox{r}_i$ is the
position of atom $i$ (including displacement).
 The other electronic matrix elements are:
 Cu-site ($U_d$) and O-site ($U_p$) repulsions
 for $U_{i}$, and the nearest-neighbor Cu-O repulsion ($U_{pd}$).
We define $\Delta=E_p-E_d + U_{pd}$ and $\epsilon=2 (E_p-E_d)+ U_p$.
In the following we adopt the notation that Cu sites are labeled
with $\bbox{i}$ and O sites  or the corresponding Cu-Cu bond
with $\bbox{i+\delta} /2$. In this notation the position of
an O ion is given by
$\bbox{r}_{\bbox{i}+\bbox{\delta} /2}=a (\bbox{i}+\bbox{\delta} /2)
+\bbox{ u}_{\bbox{i}+\bbox{\delta} /2}$, with $a$ the lattice constant.

\subsection{Effective Charges and Dipole Moment Operator}

To calculate the coupling constants of light with
 one- and two-phonon-multimagnon processes we first obtain
a low energy  Hamiltonian  as a perturbation expansion valid
when  $t<<\Delta,\epsilon,U_d$ and when the phonon field and
the electric field vary slowly with respect to typical gap frequencies,
\begin{equation}
\label{h}
H=\sum_{\bbox{i},\bbox{\delta}}
 J(\bbox{ E},\{\bbox{ u}_{\bbox{i}+\bbox{\delta} /2}\})
B_{\bbox{i}+\bbox{\delta} /2}+H_{\rm ph}-\bbox{ E} \bbox{ P}_{\rm ph} .
\end{equation}
Here $B_{\bbox{i}+\bbox{\delta} /2}=\bbox{S_iS_{i+\delta}}$ with
$\bbox{S_i}$ spin operators, $H_{\rm ph}$ is
the phonon Hamiltonian containing spring constants and masses for
 the O ions ($M$) and $\bbox{ P}_{\rm ph}$.
 The first term in Eq.~(\ref{h}) contains the
spin-dependent fourth order correction in $t$ whereas fourth,
second and zero order spin-independent processes
are collected in the last two terms.
To compute $J$ we can use the three center system
 Cu$_{\rm L}$-O-Cu$_{\rm R}$ of Fig.~\ref{con}.
\begin{figure}
See Ref.~\cite{lor95}
\vskip 7truecm

\caption{Schematic representation of the cluster used in the calculations.
Full dots represent Cu's and open dots O's. Thick arrows represent the spin,
thin short arrows represent lattice displacements and thin long arrows
represent the direction of the electric field. We have represented
${\bf u}_0$ in configuration A. In general its direction is equal to the
direction of the electric field.}
\label{con}
\end{figure}
In order to generalize our result for arbitrary spin $S$
we define the saturated ferromagnetic state
$$|{\rm F}\rangle = |m_{\rm L}=S,S_{\rm L}=S;m_{\rm R}=S,S_{\rm R}=S\rangle$$
and the N\'eel state,
$$|{\rm N}\rangle =|m_{\rm L}=S,S_{\rm L}=S;m_{\rm R}=-S,S_{\rm R}=S\rangle.$$
Here $m_{\rm R,L}$ is the $z$ component and $S_{\rm L,R}$ is the
magnitude of the total spin of the ion. The superexchange can be calculated as,
\begin{equation}
J=\frac1{S^2} (\langle {\rm F}|VRVRVRV|{\rm F}\rangle -
\langle {\rm N}|VRVRVRV|{\rm N}\rangle )
\end{equation}
where $R=(1-P)/(E_0-H_0)$, $P$ projects on the manifold with
$S_{\rm L}=S_{\rm R}=S$, $H_0$
 contains all site-diagonal terms  of the Hamiltonian Eq.~(\ref{ham})
and $V$ the $d-p$ hybridization terms. For details see Ref.~\cite{esk93}.
In the case of a 3-d S=1 system like NiO we can orient the $z$ axis
in the direction of the Ni-O-Ni bond, then $t$ should be taken as the
hybridization between $d_{3z^2-r^2}$ orbitals and the $p_z$  orbital and
is given by $pd\sigma$ in terms of Slater and Koster integrals\cite{sla54}.
Also appropriate values of $U_d$ should be used\cite{zaa87}.
The same is valid for La$_2$NiO$_4$ if the crystal field splitting
between the $d_{3z^2-r^2}$ and the $d_{x^2-y^2}$ orbitals is neglected.

 We only need to consider the three configurations (A,B,C) of the L-R bond
and the electric field. Next we Taylor expand
$J$ to first order in $\bbox{E}$ and second order in
$\{\bbox{ u}_{\bbox{i}+\bbox{\delta} /2}\}$,
\begin{eqnarray}
\label{j}
J&=&J_0+\eta (u_{\rm L} -  u_{\rm R}) - E [q_{\rm I} u_0 +
\lambda q_{\rm A} (2 u_0 -u_{\rm L} - u_{\rm R}) ]
\nonumber \\
&-& E [ \xi_{\rm I} u_0 (u_{\rm L} - u_{\rm R} )  + \lambda
\xi_{A} (2 u_0 -u_{\rm L} - u_{\rm R}) ( u_{\rm L} - u_{\rm R})] \nonumber \\
&+& ...
\end{eqnarray}
Here $\lambda=1$ for configuration A and  $\lambda=0$ for configurations
 B and C. In each configuration the displacement of the central O and the
electric field are parallel, i.e.
$\bbox{ E}=E \bbox{\hat{e}},  \bbox{ u}_0=u_0 \bbox{\hat{e}}$.
The direction of $ \bbox{\hat{e}}$ is the same as the arrows at the bottom of
Fig.~\ref{con}. $u_{\rm L}$ and $u_{\rm R}$ are only relevant in
 configuration A.
 $u_{\rm L}=   u_{\rm L1} + u_{\rm L2} - u_{\rm L3} $,
 $u_{\rm R}= - u_{\rm R1} + u_{\rm R2} + u_{\rm R3}$.
The numbering
and the direction of the displacements are shown in Fig.~\ref{con}.
Other terms quadratic in $u$'s have been neglected, they
renormalize the spring constants in  $H_{\rm ph}$ and give
the magnon-two-phonon interaction.
The first term in Eq.~(\ref{j}) is the superexchange in absence of
the electric and phonon fields,
\begin{equation}
J_0 =
\frac{t^4}{S^2\Delta^2} [\frac{1}{U_d} + \frac{2}{\epsilon}].
\end{equation}
The remaining quantities are a magnon-phonon
coupling constant,
\begin{equation}
\eta=   \frac{-t^4\beta}{S^2\Delta^2} [\frac{1}{\Delta}(\frac{1}{U_d} +
\frac{2}{\epsilon})+ \frac{2}{\epsilon^2}],
\end{equation}
   effective charges associated with
{\em one} phonon and multimagnon processes,
\begin{eqnarray}
q_{\rm I}&=& -e  \frac{2t^4}{S^2\Delta^2} [\frac{1}{\Delta}(\frac{1}{U_d} +
\frac{2}{\epsilon})+ \frac{2}{\epsilon^2}],\\
q_{\rm A}&=& -e  \frac{t^4}{S^2\Delta^2} \beta a_{pd} [\frac{2}{\Delta^2}
(\frac{1}{U_d} + \frac{2}{\epsilon})+\frac{1}{U_d}
(\frac{1}{\Delta}+\frac{2}{U_d})^2],
\end{eqnarray}
and   effective charges associated with
{\em two}-phonon and multimagnon processes.
\begin{eqnarray}
\label{iso}
\xi_{\rm I}&=& e \frac{t^4\beta}{S^2\Delta^2} [\frac{3}{\Delta^2}
(\frac{1}{U_d} + \frac{2}{\epsilon})+\frac{8}{\epsilon^2}
(\frac{1}{\Delta}+\frac{1}{\epsilon})^2]\nonumber\\
\\
\xi_{\rm A}&=& -e \frac{2t^4\beta^2 a_{pd} }{S^2\Delta^3}(\frac{2}{U_d^3}+
\frac{3}{\Delta U_d^2} +
 \frac{4}{\epsilon\Delta^2}+\frac{2}{\epsilon^2\Delta}+\frac{3}{U_d\Delta^2})
\nonumber
\end{eqnarray}
$a_{pd}$ is the Cu-O distance, $a/2$. Within a point charge estimation the
parameter $\beta a_{pd} \approx 2 U_{pd} $.
The dipole moment is obtained  from Eq.~(\ref{h}) as
$\bbox{ P}=-\frac{\partial H}{\partial \bbox{ E}}$ and using Eq.~(\ref{j}) in
the  relevant configurations.
We get up to fourth order in $t$,
\begin{equation}
\bbox{ P} = \bbox{ P}_{\rm 1ph} +\bbox{ P}_{\rm 2ph} +\bbox{ P}_{\rm 1ph+mag} +
\bbox{ P}_{\rm 2ph+mag}
\end{equation}
The first two terms describe conventional one and two phonon absorption
processes.

 We define
$\delta B_{\bbox{i}+\bbox{\delta} /2}=B_{\bbox{i}+\bbox{\delta} /2} -
\langle B_{\bbox{i}+\bbox{\delta} /2} \rangle $ and its Fourier transform,
\begin{equation}
\label{ope}
\delta B^{\delta}_{\bbox{ p}}=\frac1N \sum_{\bbox{i}} e^{i\bbox{p i}}
\delta B_{\bbox{i}+\bbox{\delta} /2},
\end{equation}
 The lattice spacing $a$ and $\hbar$ are  set to 1.
The supraindex labels the direction  of the vector $\bbox{\delta}$
and $N$ is the number of unit cells. In the same way  the Fourier transform of
$\bbox{ u}_{\bbox{i}+\bbox{\delta}/2 }$ is given by
\begin{equation}
\bbox{ u}^{\delta}_{\bbox{ p}}
=\frac1N \sum_{\bbox{i}} e^{i\bbox{p i}}
\bbox{ u}_{\bbox{i}+\bbox{\delta}/2 }.
\end{equation}
After Fourier transforming, the dipole moment for one phonon and
multimagnon processes for an in-plane field in the $x$ direction
is,
\begin{eqnarray}
\label{dip}
P^{x}_{\rm 1ph+mag}&=& N[q_{\rm I} \sum_{\bbox{ p}\delta}
 \delta B^{\delta}_{-\bbox{ p}}
u^{\delta}_{x\bbox{ p}}\nonumber\\
+\lambda 4  &q_{\rm A}& \sum_{\bbox{ p}\delta} \sin(\frac{p_x}{2})
\sin(\frac{p_\delta}{2}) \delta B^x_{-\bbox{ p}} u_{\delta\bbox{ p}}^{\delta}],
\end{eqnarray}
and $\lambda=1$.
For an electric field perpendicular to the plane we have,
\begin{equation}
P^{z}_{\rm 1ph+mag}= q_{\rm I} \sum_{\bbox{ p}\delta}
\delta B^{\delta}_{-\bbox{ p}} u^{\delta}_{z\bbox{ p}}.
\end{equation}
In a 3-d system like NiO, $\lambda=1$ and we get the analogous of
Eq.~(\ref{dip}) for the three directions and the sum over $\delta$
is also for the  three directions.

The  term proportional to $q_{\rm I}$
 is isotropic being present in any configuration.
Looking at the cluster in Fig.~\ref{con} it can be
understood as a spin dependent correction to the charge
on O$_0$.
 Its physical origin is that fourth order corrections to the charges
 involve spin dependent processes.
For example if the  spins in Cu$_{\rm L}$ and Cu$_{\rm R}$ are parallel they
cannot both transfer to  O$_0$ whereas if they are antiparallel they can.
 Fig.~\ref{pro}(a)
illustrates a typical process efficient in  configuration B.

 An alternative way of deriving $q_{\rm I}$ is with the aid of the
 Hellmann-Feynman theorem. We can add to the electronic Hamiltonian
of the  Cu$_{\rm L}$-O-Cu$_{\rm R}$
a term $\varepsilon n_{\rm O}$ where
$n_{\rm O}=n_{\rm O\uparrow}+n_{\rm O\downarrow}$ and
$n_{\rm O\sigma}$  is the occupation number operator for the central
O ion and  spin $\sigma$. Now we can calculate the expectation
values of  the resulting Hamiltonian $H(\varepsilon)$
 in the ground state $|\varepsilon\rangle$. We have that the
{\em total} charge on the O is given by the following derivative of the
energy,
\begin{equation}
\langle 0|n_{\rm O}|0\rangle
=e \frac{\partial \langle \varepsilon
|H(\varepsilon)|\varepsilon\rangle}{\partial \varepsilon}
|_{\varepsilon =0}.
\end{equation}
  Since the
term proportional to $J_0(\varepsilon)$ in $\langle \varepsilon
|H(\varepsilon)|\varepsilon\rangle$
is the magnetic part of the energy its derivative is related to
 the magnetic dependent part of the O charge,
\begin{equation}
q_{\rm I}=e \frac{\partial J_0(\varepsilon)}{\partial
\varepsilon}|_{\varepsilon =0},
\end{equation}
and since $\varepsilon$ just renormalizes $\Delta$ we have formally
\begin{equation}
q_{\rm I}=e \frac{\partial J_0}{\partial \Delta}.
\end{equation}
This expression is useful to obtain $q_{\rm I}$ if one can get $J_0$
from a different method.
\begin{figure}
See Ref.~\cite{lor95}
\vskip 9truecm
\caption{Typical processes contributing to the isotropic (a) and anisotropic
(b) effective charges. The meaning of the symbols is the same as in
Fig.~2. }
\label{pro}
\end{figure}
The term proportional to $q_{\rm A}$ in Eq.~(\ref{dip})
is anisotropic being present for an in-plane field only
or when there are Cu-Cu bonds oriented in the direction of the field.
 It originates from a ``charged phonon'' like effect\cite{ric79}.
Consider the configuration in which the electric field and the displacement
of O$_0$ are both parallel to the Cu$_{\rm L}$-Cu$_{\rm R}$ bond
 (A in Fig.~\ref{con}),
and a phonon in which the O's around Cu$_{\rm L}$ breathe in and the
O's around Cu$_{\rm R}$ breathe out. (We don't need to consider zero momentum
phonons to couple to light since it is the total momentum,
magnons plus phonons  which
 has to add to zero.) The Madelung potential in  Cu$_{\rm R}$ decreases, and
in Cu$_{\rm L}$ it increases, creating a displacement of charge from left
to right that
contributes to the dipole moment.  Fig.~\ref{pro}(b) illustrates a typical
process. But again this effect is spin dependent since if the two spins
are parallel they cannot both transfer to Cu$_{\rm R}$ and hence
one gets a spin dependent correction to the dipole charge. In a highly
covalent material like the cuprates one expect the charge phonon effects
to be quite strong and dominate the effective charges (see Sec.~\ref{ani}).

\subsection{Optical absorption}

 The real part of the
 optical conductivity due to the processes described in the
previous sections  is given by
the dipole-moment-dipole-moment correlation function. In Zubarev's
notation\cite{zub60}
\begin{equation}
\sigma = -\frac{2\pi\omega}{N V_{\rm Cu}}
 {\rm Im} (\langle\!\langle  P_{\rm 1ph+mag} ;
P_{\rm 1ph+mag} \rangle\!\rangle)
\end{equation}
Here $V_{\rm Cu}$ is the volume associated with a Cu ion, i.e.
$N V_{\rm Cu} $ is the total volume of the system per Cu-O layer.
Now we assume for simplicity  that only the $u^{\delta}_{e\bbox{ p}}$'s
with the same $\delta$ and $e$ mix. To zero order in the magnon-phonon
interaction we can decouple the magnetic system from the phonon system.
In this approximation the eigenstates of the system are products of
phonon states times magnetic states and by writing the
 Lehmann representation of the Green function one can factor out
all phonon matrix elements. Using that
$$\langle 0_{\rm ph} |u^{\delta}_{e\bbox{ p}}
u^{\delta}_{e\bbox{-p}} |0_{\rm ph}  \rangle=
\frac{1}{2M\omega^{\delta}_{e\bbox{ p}}N } $$
 with
$|0_{\rm ph} \rangle $ the phonon vacuum and $\omega^{\delta}_{e\bbox{
p}}$
the phonon frequency. We get,
\begin{eqnarray}
\label{sdw}
&\sigma& = -\frac{\pi\omega}{M V_{\rm Cu}} \sum_{\bbox{ p}} {\rm Im}\Bigl[
\frac{q_{\rm I}^2}{\omega_{\perp\bbox{ p}}}
 \langle\!\langle \delta B^y_{\bbox{ -p}};
\delta B^y_{\bbox{ p}}\rangle\!\rangle^{\omega_{\perp\bbox{ p}}} \nonumber \\
&+&  \frac{16 \lambda^2 q_{\rm A}^2\sin^2(\frac{p_x}{2})\sin^2(\frac{p_y}{2})+
(4\lambda q_{\rm A} \sin^2(\frac{p_x}{2})-q_{\rm I})^2}
{\omega_{\parallel\bbox{ p}}} \\
& & \langle\!\langle \delta B^x_{\bbox{ -p}};
\delta B^x_{\bbox{ p}}\rangle\!\rangle^{\omega_{\parallel\bbox{ p}}}
 \Bigl] .\nonumber
\end{eqnarray}
Here $\omega_{\parallel\bbox{ p}}$ is the frequency of the
$ u_{x\bbox{ p}}^x$ and $u^y_{y\bbox{ p}}$ phonons and
 $\omega_{\perp\bbox{ p}}$ is the frequency of the
$ u^y_{x\bbox{ p}}$ and $u^x_{y\bbox{ p}}$ phonons.
$\omega_{\parallel\bbox{ p}}$ can be associated with the frequency of
Cu-O stretching mode phonons and $\omega_{\perp\bbox{ p}}$ with that of
Cu-O bending mode phonons.
The supraindex in the Green functions indicates that the poles
should be shifted by that amount.

\section{THE TWO MAGNON PROBLEM IN INTERACTING SPIN WAVE THEORY}
\label{lin}
To compute the magnon-magnon Green functions we use
interacting spin-wave theory\cite{man91} with a Holstein-Primakoff
transformation. On the A  sublattice we put
\begin{mathletters}
\begin{eqnarray}
\label{hpa}
&S^+_{\bbox{i}}& =
\sqrt{2S(1-\frac{b_{\bbox{i}}^{\dag} b_{\bbox{i}}}{2S}) } b_{\bbox{i}}
 \nonumber\\
&S^-_{\bbox{i}}& = b_{\bbox{i}}^{\dag}
\sqrt{2S(1-\frac{b_{\bbox{i}}^{\dag} b_{\bbox{i}}}{2S})}\\
&S^z_{\bbox{i}}&=S-b_{\bbox{i}}^{\dag} b_{\bbox{i}} \nonumber
\end{eqnarray}
and on the B sublattice,
\begin{eqnarray}
\label{hpb}
&S^+_{\bbox{i}}& =
b_{\bbox{i}}^{\dag}\sqrt{2S(1-\frac{b_{\bbox{i}}^{\dag}
b_{\bbox{i}}}{2S})}\nonumber\\
&S^-_{\bbox{i}}& =
\sqrt{2S(1-\frac{b_{\bbox{i}}^{\dag} b_{\bbox{i}}}{2S})}b_{\bbox{i}}\\
&S^z_{\bbox{i}}&=-S+b_{\bbox{i}}^{\dag} b_{\bbox{i}} \nonumber
\end{eqnarray}
\end{mathletters}
where $b_{\bbox{i}}$  is  a Boson operator.

Now the Hamiltonian can be formally expanded in powers of $1/S$. From
now on we adopt a more conventional notation and  drop the $0$
subindex in $J_0$. We define $n_i= b_{\bbox{i}}^{\dag} b_{\bbox{i}}$.
The Hamiltonian reads $H= E_{\rm N\acute{e}el} + H_0 + H_1$. With,
\begin{eqnarray}
 &E&_{\rm N\acute{e}el}= -\frac12 JS^2zN  \\
 &H&_0 =  S J z \sum_{\bbox{i}}n_{\bbox{i}} + SJ \sum_{<\bbox{ij}>}
  (b_{\bbox{i}}^{\dag} b_{\bbox{j}}^{\dag}+ h.c.)\\
 &H&_1 = - J \sum_{<\bbox{ij}>} n_{\bbox{i}}n_{\bbox{j}} - \frac{J}4
\sum_{<\bbox{ij}>}
[b_{\bbox{i}}^{\dag} b_{\bbox{j}}^{\dag} (n_{\bbox{i}}+n_{\bbox{j}}) + h.c.]
\end{eqnarray}
$<\bbox{ij}>$ indicates that nearest neighbor pairs are counted once in
 the sum,
$z$ is the coordination number,
 $E_{\rm N\acute{e}el}$ is the classical N\'eel energy, $H_0$ is
the linear spin wave Hamiltonian and $H_1$
is the spin-wave-spin-wave  interaction. Notice that
the Hamiltonian is  invariant
under the exchange of the sublattices and so we don't need to distinguish
between them. Accordingly we work in the non-magnetic Brillouin zone.

  The non-interacting part, $H_0$ is diagonalized
by the Bogoliubov transformation,
\begin{eqnarray}
Q_{\bbox{k}}^{\dag}&=&
u_{\bbox{k}}b^{\dag}_{\bbox{k}}-v_{\bbox{k}}b_{\bbox{-k}}\nonumber\\
\label{swo}& &\\
Q_{\bbox{k}}&=&
u_{\bbox{k}}b_{\bbox{k}}-v_{\bbox{k}}b_{\bbox{-k}}^{\dag}\nonumber
\end{eqnarray}
 where $b_{\bbox{k}}$ is the Fourier transform of $b_{\bbox{i}}$ and
\begin{eqnarray}
u_{\bbox{k}}&=&\sqrt{\frac{1+\omega_{\bbox{k} }}
 {2\omega_{\bbox{k}}}},\nonumber\\
& &\\
v_{\bbox{k}}&=&-{\rm sig}(\gamma_{\bbox{k}})
\sqrt{\frac{1-\omega_{\bbox{k}}}{2\omega_{\bbox{k}}}},\nonumber
\end{eqnarray}
and
\begin{eqnarray}
\gamma_{\bbox{k}}&=&\frac2z \sum_{\delta}\cos(k_{\delta}),\\
\omega_{\bbox{k}}&=&\sqrt{1-\gamma_{\bbox{k}}^2}.
\end{eqnarray}
Now we can normal order the Hamiltonian with respect to the non-interacting
spin-wave ground state. This is equivalent to writing  the Hamiltonian
as a Hartree-Fock part plus fluctuations which we latter treat in the RPA.
After normal ordering the  Hamiltonian can be written as,
\begin{equation}
\label{hsw}
H=E_{\rm N\acute{e}el}+E_{\rm SW}+
\sum_{\bbox{k}}E_{\bbox{k}}Q_{\bbox{k}}^{\dag} Q_{\bbox{k}}+ V_{\rm res},
\end{equation}
where
 $E_{\rm SW}= \frac14 JSNz\zeta(1+\zeta/2S)$,
$E_{\bbox{k}}=E_{\rm m}\omega_{\bbox{k}} $,
$E_{\rm m}=zSJ(1+\zeta/2S)$,
 $\zeta$ is the Oguchi correction,
\begin{equation}
\zeta=1-\frac1N \sum_{\bbox{k}} \omega_{\bbox{k}}.
\end{equation}
 For $z=4$  as in the layered materials,  $\zeta\approx 0.158$.
$V_{\rm res}=V_{\rm res}^{\parallel}+V_{\rm res}^{\perp}$
 contains the normal ordered product of
the interacting part i.e.
\begin{eqnarray}
V_{\rm res}^{\parallel}&=& -J \sum_{<\bbox{ij}>} :n_{\bbox{i}}n_{\bbox{j}}: \\
V_{\rm res}^{\perp}&=&- \frac{J}4 \sum_{<\bbox{ij}>}
 :[b_{\bbox{i}}^{\dag} b_{\bbox{j}}^{\dag}(n_{\bbox{i}}+n_{\bbox{j}}) + h.c.]:
\end{eqnarray}
The first term originates from the Ising part of the interaction.
In the Ising limit it is easy to check that the effect of this term is
to shift the energy of two nearest-neighbors  spin flips
from the noninteracting value $4J$ to $3J$. In the  Heisenberg limit
and in the case of the Raman line shape as an example, the non interacting
line shape has a peak at $4J$ and this term shifts it to close to $3J$.
This term is the more important one to get the
correct line shape. The second term is a correction to the
 exchange due to the kinematic interaction.

The next step is to
  put $V_{\rm res}$ in terms of the spin-wave operators Eq.~(\ref{swo}).
 Then we  evaluate the normal order and we
do the RPA approximation. This consists in this case, in keeping
 only those  terms in $V_{\rm res}$ which create or destroy a pair of magnons.
With this we get,
\begin{equation}
\label{vres}
V_{\rm res }^{\rm RPA} =\frac1{2N}\sum_{1234}\delta(1+2-3-4)
\Gamma_{1234}Q_1^{\dag}Q_2^{\dag}Q_3Q_4\\
\end{equation}
$1, 2,...$ stands for $\bbox{k}_1, \bbox{k}_2,...$ and the vertex are
given in the Appendix~\ref{ver.a}.

We define
\begin{equation}
 g_{\bbox{p}\bbox{q}_1\bbox{q}_2}=
\langle\!\langle Q_{\frac12{\bbox p}+{\bbox q}_1}
 Q_{ \frac12{\bbox p}-{\bbox q}_1}
;Q_{\frac12{\bbox p}+{\bbox q}_2}^{\dag}
 Q_{\frac12{\bbox p}-{\bbox q}_2}^{\dag}
\rangle\!\rangle.
\end{equation}
$\bbox{p}$ is the total momentum of a magnon pair and ${\bbox q}_1$,
${\bbox q}_2$, the relative momentum.
In the following we make indiscriminate use of the property that
$ g_{\bbox{p}\bbox{q}_1\bbox{q}_2}$ does not change when
 $\bbox{q}_1\rightarrow-\bbox{q}_1$.

It is convenient for latter use to define the following
auxiliary functions
\begin{mathletters}
\label{albe}
\begin{eqnarray}
\label{alp}
&\alpha&^{\pm}_{\bbox{p}\bbox{q}}=
u_{\frac12{\bbox p}+{\bbox q}}u_{ \frac12{\bbox p}-{\bbox q}}\pm
v_{\frac12{\bbox p}+{\bbox q}}v_{ \frac12{\bbox p}-{\bbox q}},\\
&\beta&^{\pm}_{\bbox{p}\bbox{q}}=
u_{\frac12{\bbox p}+{\bbox q}}v_{ \frac12{\bbox p}-{\bbox q}}\pm
v_{\frac12{\bbox p}+{\bbox q}}u_{ \frac12{\bbox p}-{\bbox q}},
\end{eqnarray}
\end{mathletters}
 the following ``different sublattice'' form factors,
\begin{eqnarray}
\label{fld}
&f&^{1\delta}_{\bbox{p}\bbox{q}}=\alpha^{+}_{\bbox{p}\bbox{q}}
\cos(q_{\delta}),\nonumber\\
&f&^{2\delta}_{\bbox{p}\bbox{q}}=\alpha^{-}_{\bbox{p}\bbox{q}}
\cos(q_{\delta}),\\
&f&^{3}_{\bbox{p}\bbox{q}}=\beta^+_{\bbox{p}\bbox{q}},\nonumber
\end{eqnarray}
and ``same sublattice'' form factors,
\begin{eqnarray}
\label{hld}
&h&^{1\delta}_{\bbox{p}\bbox{q}}=
\beta^+_{\bbox{p}\bbox{q}}\cos(q_{\delta}),\nonumber\\
&h&^{2\delta}_{\bbox{p}\bbox{q}}=
\beta^-_{\bbox{p}\bbox{q}}\sin(q_{\delta}),\\
&h&^{3}_{\bbox{p}\bbox{q}}=\alpha^+_{\bbox{p}\bbox{q}},\nonumber\\
&h&^{4}_{\bbox{p}\bbox{q}}=\alpha^-_{\bbox{p}\bbox{q}}.\nonumber
\end{eqnarray}
The $f$'s  have the property that
$f_{\bbox{p}\bbox{q+\pi}}=-f_{\bbox{p}\bbox{q}}$, with
$\bbox{\pi}=(\pi,\pi,...)$. If
we Fourier transform  them in $\bbox{q}$ we see that they are different
from zero  in  different sublattices. For the $h$'s,
$h_{\bbox{p}\bbox{q+\pi}}=h_{\bbox{p}\bbox{q}}$
and they are different from zero on the same sublattice.
They will  allow  us to classify the Green functions in ``different
sublattices'' Green functions and ``same sublattice'' Green function
where the relevant coordinate is the  distance between
the spin operators. For example in the case of the operator Eq.~(\ref{ope})
since $\bbox{\delta}$ joints different sublattices we need the
former.

By replacing $V_{\rm res}$ by $V_{\rm res}^{\rm RPA}$
in the Hamiltonian Eq.~(\ref{hsw}) the RPA
equation of motion for the Green functions can
be computed by standard methods\cite{zub60}. No further approximations
are needed to obtain the Eqs.~(\ref{gsa}),(\ref{gdi}).  We get
\begin{equation}
\label{gsw}
g_{\bbox{p}\bbox{q}_1\bbox{q}_2} =
\frac{\delta_{\bbox{q}_1,\bbox{q}_2}+\delta_{\bbox{q}_1,-\bbox{q}_2}}{2\pi}
g^0_{\bbox{p}\bbox{q}_1}+\frac1N
\sum_{\bbox{q}} g^0_{\bbox{p}\bbox{q}_1} \Gamma_{\bbox{p}\bbox{q}_1\bbox{q}}
g_{\bbox{p}\bbox{q}\bbox{q}_2},
\end{equation}
where $g^0_{\bbox{p},\bbox{q}_1}=(\omega- E_{\frac12{\bbox p}+{\bbox q}_1}
-E_{ \frac12{\bbox p}-{\bbox q}_1})^{-1}$ and the vertex is given
in the Appendix~\ref{ver.a}.
Since the vertex is a sum of separable potentials the
equations can be solved in integral  form.
We define same sublattice Green functions,
\begin{eqnarray}
\label{fi0}
&K^{(0)}_{\mu\nu}&=\frac1N \sum_{\bbox{q}}
 g_{\bbox{p}\bbox{q}}^0 h^{\mu }_{\bbox{pq}}
h^{\nu}_{\bbox{pq}}\\
&K_{\mu\nu}&=\frac{\pi}{N} \sum_{\bbox{q}\bbox{q}'}
 g_{\bbox{p}\bbox{q}\bbox{q}'}
 h^{\mu}_{\bbox{pq}} h^{\nu}_{ \bbox{pq}'}
\end{eqnarray}
where $\mu,\nu=l,\delta $ for $l=1,2$ and  $\mu,\nu=l$ for $l=3,4$.
Analogously the different sublattice Green functions are given by,
\begin{eqnarray}
\label{gi0}
&G^{(0)}_{\mu\nu}&=\frac1N \sum_{\bbox{q}}
 g_{\bbox{p}\bbox{q}}^0 f^{\mu }_{\bbox{pq}}
f^{\nu}_{\bbox{pq}}\\
\label{gmn}
&G_{\mu\nu}&=\frac\pi{N} \sum_{\bbox{q}\bbox{q}'}
 g_{\bbox{p}\bbox{q}\bbox{q}'}
 f^{\mu}_{\bbox{pq}} f^{\nu}_{ \bbox{pq}'}
\end{eqnarray}
 It is easy to see that  similar zero order Green functions
with an $f$ and an $h$ factor in the kernel vanish.
Since the vertex [Eq.~(\ref{ver})] do not mix same sublattice with
different sublattice Green functions the equations
for the interacting Green functions separate in two blocks.
{}From Eq.~(\ref{gsw}) we get the following for same sublattice
Green functions,
\begin{eqnarray}
\label{gsa}
 K_{\mu\nu} =K_{\mu\nu}^{(0)}-J&\{&
\sum_{\delta} [K^{(0)}_{\mu1\delta}
+\frac12 \cos(\frac{p_{\delta}}2)
 K^{(0)}_{\mu 3}] K_{1\delta\nu}\nonumber\\
&+&\sum_{\delta} [K^{(0)}_{\mu 2\delta}
+\frac12 \sin(\frac{p_{\delta}}2)
 K^{(0)}_{\mu 4}] K_{2\delta \nu}\nonumber\\
&+&\frac12 [\sum_{\delta}  \cos(\frac{p_{\delta}}2)
 K^{(0)}_{\mu 1\delta}]  K_{3 \nu} \nonumber\\
&+&\frac12 [\sum_{\delta}
\sin(\frac{p_{\delta}}2)
 K^{(0)}_{\mu 2\delta}]  K_{4 \nu}\},
\end{eqnarray}
and for different sublattices,
\begin{eqnarray}
\label{gdi}
G_{\mu\nu} =&G_{\mu\nu}^{(0)}&-J
\{\sum_{\delta}[G^{(0)}_{\mu 1\delta}+\cos(\frac{p_{\delta}}2)
 G^{(0)}_{\mu 3}] G_{1\delta \nu}\nonumber\\
&+&\sum_{\delta} G^{(0)}_{\mu 2\delta}G_{2\delta \nu}\nonumber\\
&+& [\frac{z}2 \gamma_{\bbox{p}} G^{(0)}_{\mu 3}
+\frac12 \sum_{\delta} \cos(\frac{p_{\delta}}2)
 G^{(0)}_{\mu 1\delta}]  G_{3 \nu}\}.
\end{eqnarray}
Note that the Green functions depend on the
 total momentum $\bbox{p}$ which appears also as
a parameter in the equations.

We can expand also our two magnon Green function in Eq.~(\ref{sdw}) in
powers of  $1/S$ to lowest order we get,
\begin{eqnarray}
\label{bbo}
& &
\langle\!\langle\delta B^x_{\bbox{ -p}};\delta B^x_{\bbox{ p}}\rangle\!\rangle=
\frac{S^2}{N\pi} \times
\nonumber\\
& &[G_{1x 1x}+\cos(\frac{p_x}2)(G_{3 1x}+G_{1x 3})+
\cos^2(\frac{p_x}2) G_{3 3}]
\end{eqnarray}
This neglects small Oguchi type corrections to the prefactor\cite{can92}
 which slightly renormalizes the intensity scales in the results that follow.
First we will solve the equations in some limiting case in which
one can handle the equations analytically.

\subsection{2-d Raman}
\label{ram}
\begin{figure}
\epsfverbosetrue
\epsfxsize=10cm
$$
\epsfbox[18 144 592 500]{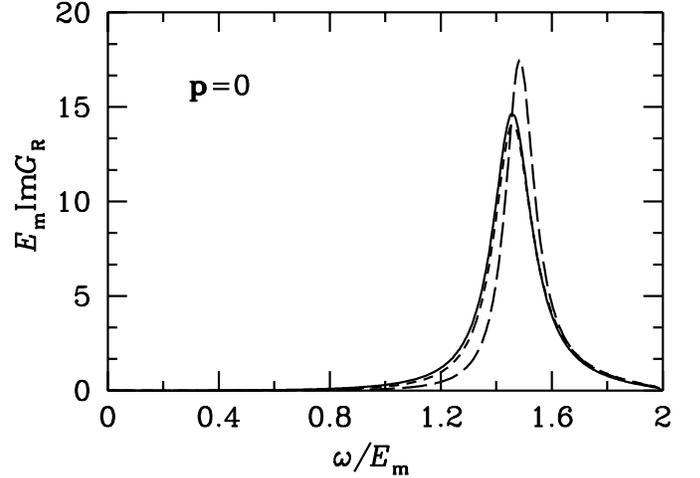}
$$
\caption{Raman line shape as given by Im$G_{\rm R}$
according to  Eq.~(\protect\ref{lra})
 (solid line), in the high energy (HE) approximation (short dashed line) and
taking  $u_{ \protect\bbox{k}}=1$, $v_{ \protect\bbox{k}}=0$
 (long dashed line). }
\label{ram.f}
\end{figure}

It is useful to see how some well known results
are recovered in this formalism. In Appendix~\ref{pxpy}
we solve the equations for the case  $p_x=p_y$.
The case of $\bbox{ p}=0$ is of particular interest
because it corresponds to the Raman line shape.
According to Ref.~\cite{par69}
for the case of a layered antiferromagnet the scattered intensity is
 proportional to the imaginary part of the following Green
function,
\begin{equation}
\frac{S^2}{\pi}G_{\rm R}\equiv
N \langle\!\langle B^x_0- B^y_0; B^x_0- B^y_0\rangle\!\rangle=
\frac{S^2}{\pi} G_{11, \bbox{ p}=0 }^d.
\end{equation}
$G^{d}_{ll'}$ is defined in the Appendix~\ref{pxpy} and,
in this case $G^{d(0)}_{11,\bbox{ p}=0}=L^{(2)}$,
 $G^{d(0)}_{22,\bbox{ p}=0}=L^{(0)}$
and $G^{d(0)}_{12,\bbox{ p}=0}=L^{(1)}$ where,
\begin{equation}
L^{(l)}=\frac{1}{N} \sum_{\bbox{ q}}
\frac{f_{\bbox{ q}}^2}{(\omega_{\bbox{ q}})^l}
\frac1{\omega-2E_{\bbox{ q}}}.
\end{equation}
and $f_{\bbox{ q}}=\cos{q_x}-\cos{q_y}$.
With this equivalences Eq.~(\ref{gd}) becomes
 equivalent to the well known expression\cite{can92},
\begin{equation}
\label{lra}
 G_{\rm R}=
\frac{L^{(2)}+\frac{J}2[L^{(0)}L^{(2)}-(L^{(1)})^2 ]}
{1+\frac{J}2(L^{(0)}+L^{(2)})+\frac{J^2}4[L^{(0)}L^{(2)}-(L^{(1)})^2 ]}.
\end{equation}
 Note that the  fact
that we use a Holstein-Primakoff representation rather than
a Dyson-Maleev  one  (as in Ref.~\cite{can92})  did not affect this result.
This is because  the effect of the exchange part
of the interaction which is different in the two formalisms
cancels out in this case due to symmetry.
In Fig.~\ref{ram.f} we show with a solid line the imaginary part of
$G_{\rm R}$  which gives the Raman line shape.
The position of the maximum is at
\begin{equation}
\label{em0}
E^{\rm max}_{(0,0)}=1.46E_{\rm m}=3.38J
\end{equation}
as is well know\cite{can92}.
Since the line shape lies at high energies  ($\omega>E_{\rm m}$)
we can get a good approximation for $G^{d(0)}_{11}$ [Eq.~(\ref{gd})]
by applying the HE approximation of Appendix~\ref{hea}.
We have  $G^{d(0)}_{++}\simeq G^{d(0)}_{22} \simeq G^{d(0)}_{12}
\simeq G^{d(0)}_{11}$, where in this case,
$G^{d(0)}_{++,\bbox{ p}=0}=\frac14 (L^{(0)}+2L^{(1)}+L^{(2)})$ and we
 get,
\begin{equation}
\label{her}
G_{\rm R}=\frac{G^{d(0)}_{++,\bbox{ p}=0}}{1+JG^{d(0)}_{++,\bbox{ p}=0}}.
\end{equation}
Here the distinction mention in Appendix~\ref{hea}
 between the HE approximation done in the vertex
alone or in the vertex and in the operator does not apply because
the operator has been explicitly constrained to have the correct symmetry.
The corresponding  line shape is  given by the short dashed line in
Fig.~\ref{ram.f} for the Raman case.
The position of the  maximum is at 1.46$E_{\rm m}$ as before
and only the intensity decrease a bit.
A  popular and more drastic approximation consists in  putting
$u_{\bbox{k}}=1$, $v_{\bbox{k}}=0$. In this case one gets
for the Raman case,
\begin{equation}
 G_{\rm R}= \frac{L^{(0)}}{1+JL^{(0)}}.
\end{equation}
This  line shape is shown with the long dashed line. The peak shifts
appreciably and is at 1.48$E_{\rm m}$. This approximation is much
worse and we will not use it in the following.

It is interesting to compare the present approach with
the pioneering work of Elliott and Thorpe\cite{ell69,ell68}.
 They  assume a N\'eel state as starting point and neglect the
 Oguchi correction.  This gives in this case\cite{par69}
 a similar line  shape but peaking at $2.71J$.
The present theory assumes a spin-wave ground state and takes
the  Oguchi correction into account. Although, at least
qualitatively a similar
result is obtain for $\bbox{ p}=0$ in the two theories,
 this changes dramatically  for  $\bbox{ p}\not= 0$.
 If one tries to extend the Elliott and Thorpe
theory for that case, nonphysical poles appear in the Green functions at
energies $E_{\frac12{\bbox p}+{\bbox q}_1}-E_{ \frac12{\bbox p}-{\bbox q}_1}$
indicating that a better ground state must be used as done here.

\subsection{IR case}
To calculate the IR line shape we need to evaluate the
Green function given by Eq.~(\ref{bbo}). The narrow
experimental line shape for the cuprates
suggests the occurrence of a sharp resonance for some values of
$\bbox{ p}$.
This can not occur close to $\bbox{ p}=(0,0)$ or $(\pi,\pi)$ because
there the continua of two magnon excitations extends
from zero to 2$E_{\rm m}$ and the imaginary part of the
non interacting  Green function is relatively large at the energies
of typical two-magnon excitations.
 This changes at different values of $\bbox{ p}$,  in particular
for $\bbox{ p}=(\pi,0)$ there is a gap in the spectrum from 0
to $E_{\rm m}$ and the  imaginary part of the non interacting
Green function is very small where the resonance occurs.
This case is analyzed in detail next.

\subsubsection{Bimagnon at $\bbox{ p}=(\pi,0)$}

\begin{figure}
\epsfxsize=10cm
$$
\epsfbox[18 144 592 500]{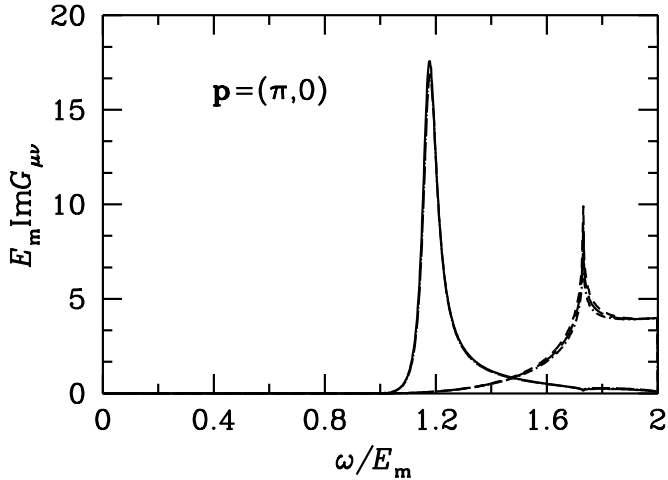}
$$
\caption{Im$G_{1x1x}$ at $\bbox{p}=(\pi,0)$
in the full theory of Eq.~(\protect\ref{gxx})  (solid line),
 and almost indistinguishable,
in the HE approximation (doted long dashed line).
We also show (very close to each other)
Im$G_{1x1x}^{(0)}$ (short dashed line),
Im$G_{1x2x}^{(0)}$ (long dashed line),  Im$G_{2x2x}^{(0)}$
(doted short dashed line).  }
\label{gpi0.f}
\end{figure}

The relevant  Green function for this problem is [Eq.~(\ref{bbo})]
\begin{equation}
\langle\!\langle\delta B^x_{-(\pi,0)};
\delta B^x_{(\pi,0)}\rangle\!\rangle=
\frac{S^2}{N\pi} G_{1x1x}.
\end{equation}
In the Appendix~\ref{px0}  we solve Eq.~(\ref{gdi}) for this momentum
in any dimension. $ G_{1x1x} $ is given by Eq.~(\ref{gxx}).
In Fig.~\ref{gpi0.f} we show with a solid line the imaginary part of this
Green function. A sharp resonance occurs indicating that
a virtual  bound state (bimagnon) is formed. The maximum is at
\begin{equation}
\label{emp}
E^{\rm max}_{(\pi,0)}=1.179E_{\rm m}=2.731J
\end{equation}
We show also Im$G_{1x1x}^{(0)}$, Im$G_{1x2x}^{(0)}$, Im$G_{2x2x}^{(0)}$
 This represents the density of states of the continua
of two magnon excitations in which  the bimagnon can decay.  As mention
before it starts at $E_{\rm m}$ and is  very small at the position
of the bimagnon pole. This explains the long lifetime of the bimagnon.
 We see that  $G^{d(0)}_{1x1x}\simeq G^{d(0)}_{2x2x}
 \simeq G^{d(0)}_{1x2x}$.
This is because at high energies $v_{\bbox{k}}$ is quite small
[see Eqs.~(\ref{alp}),(\ref{fld}),(\ref{gi0})]
and all of them are very close to $G^{(0)}_{+x+x}$ defined
in Appendix.~\ref{hea}.
 We can then apply the HE approximation  of Appendix.~\ref{hea}.
For simplicity here we use the HE approximation
both in the vertex and in the operator since
in this case there is no low energy spurious spectral weight
i.e. we neglect $G'_{+x-x},G'_{-x-x}$ in Eq.~(\ref{gp11}).
We get  the familiar RPA  like form,
\begin{equation}
\label{gpi0}
G'_{1x1x}= \frac{G^{(0)}_{+x+x}} {1+2JG^{(0)}_{+x+x}}.
\end{equation}
The imaginary part of this is shown with doted long dashed line
 in Fig.~\ref{gpi0.f}.
Again we see that the HE approximation is very accurate.
 In particular the peak does not shift and there is only a small decrease in
the intensity.

\subsubsection{Bimagnon for general $\bbox{p}$}

To obtain the total line shape we have to integrate the
 contributions from the whole Brillouin zone so we need
the Green function at all  values of $\bbox{p}$.
By the same arguments as before we can use the HE approximation
of Appendix~\ref{hea}.  We expect
this to work well because of the following reasons.
 i) We expect that the total line shape  will be  determined by the
 sharp excitations that occur close to momentum $(\pi,0)$ and
we have seen that in this region the approximation does extremely good.
ii) Far from this region the approximation is also good
 as we have already shown for the Raman case.

\begin{figure}
\epsfxsize=8cm
$$
\epsfbox[0 500 596 842]{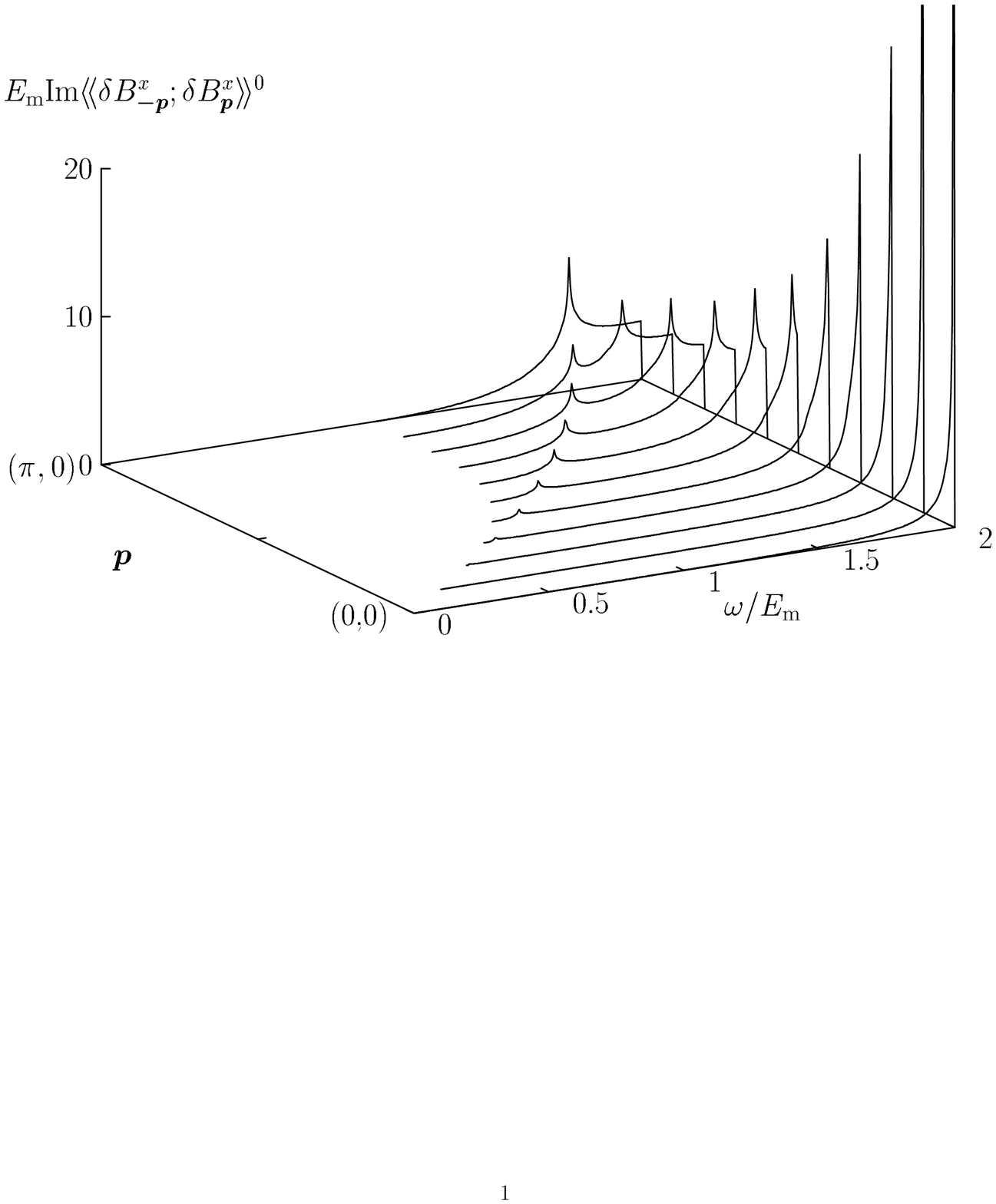}
$$
\epsfxsize=8cm
$$
\epsfbox[0 500 596 842]{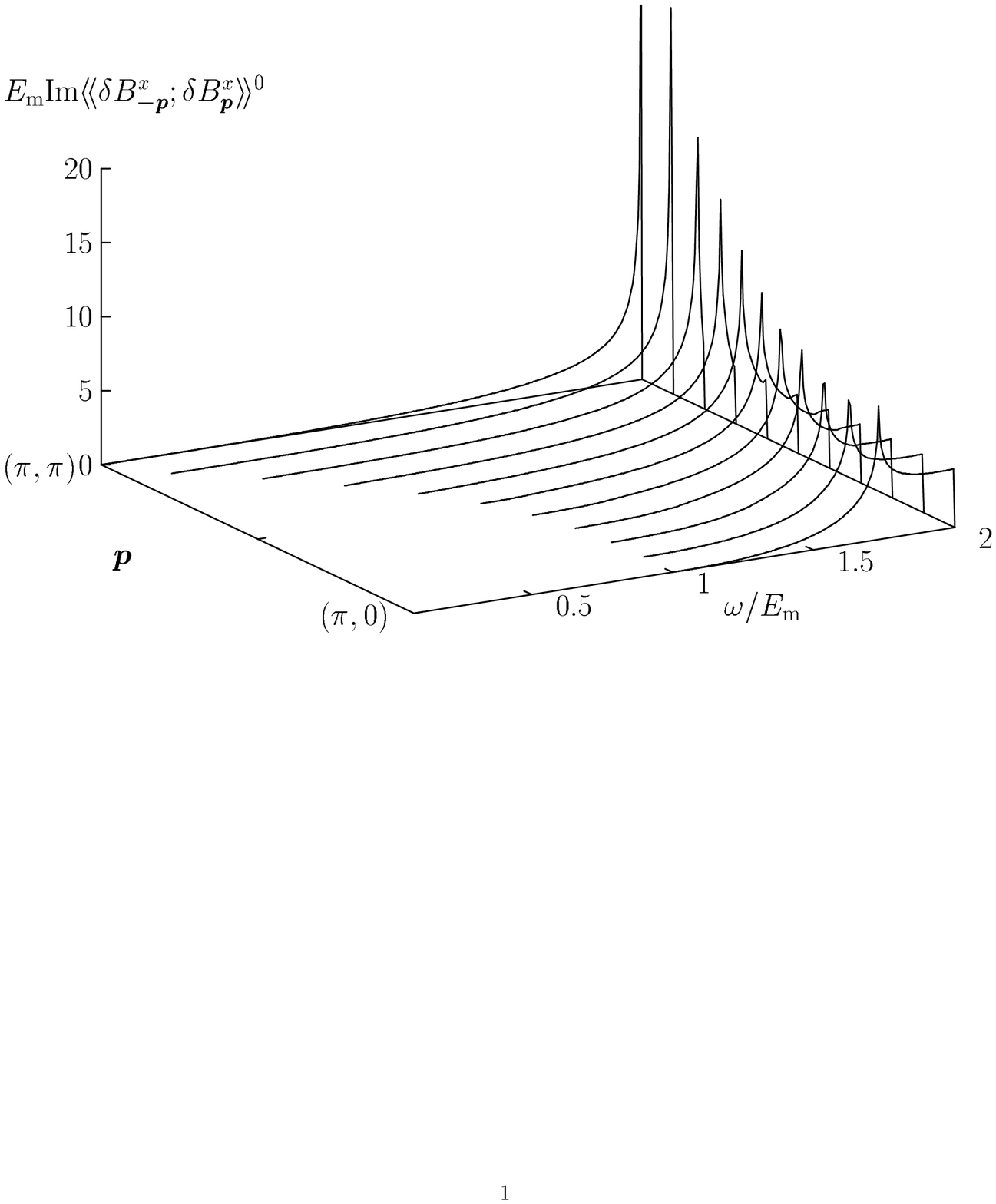}
$$
\epsfxsize=8cm
$$
\epsfbox[0 300 596 842]{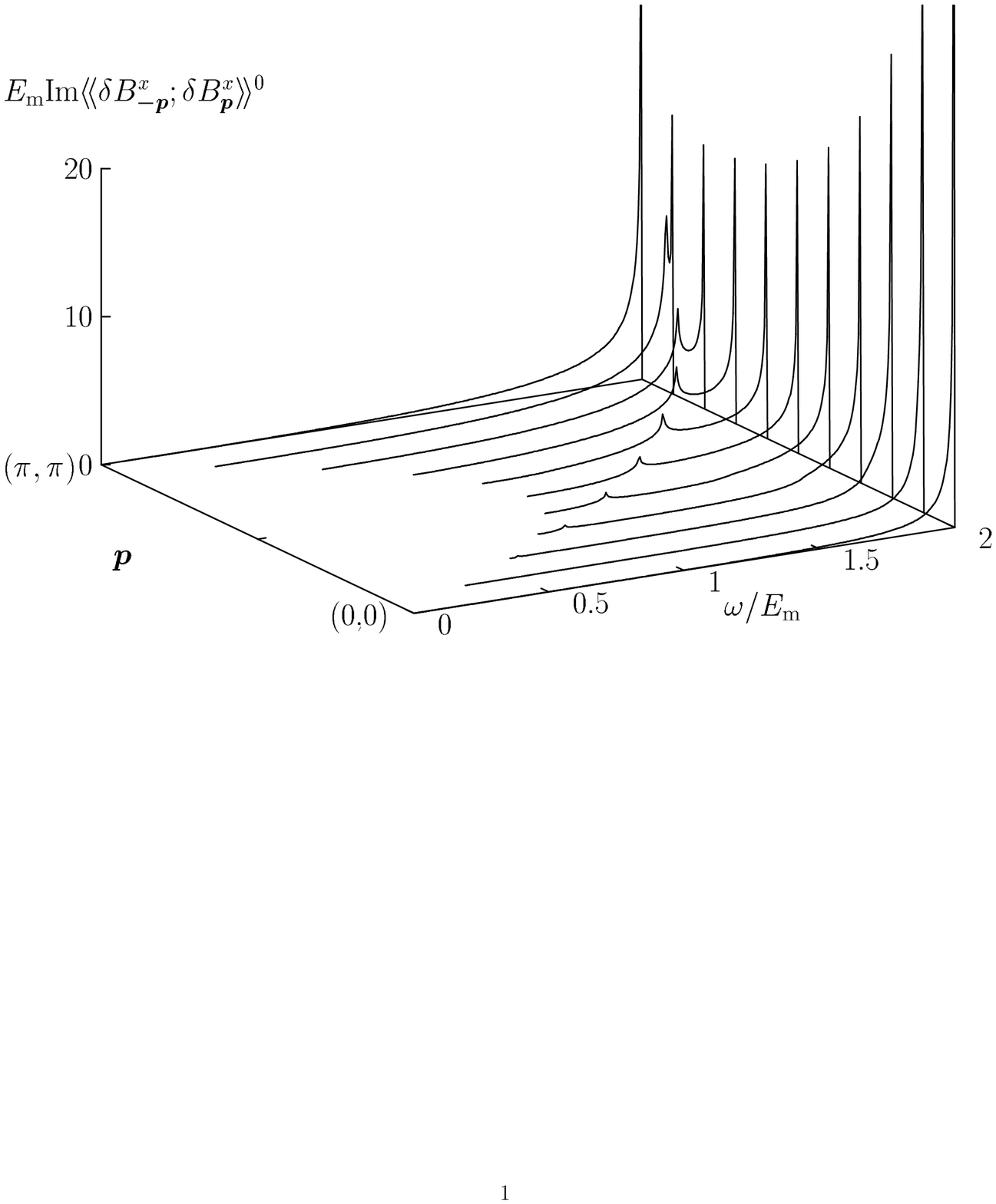}
$$
\caption{Im$\langle\!\langle\delta B^x_{\bbox{ -p}};
\delta B^x_{\bbox{ p}}\rangle\!\rangle$
 in the noninteracting  ($S=\infty$) case
as a function of  $\bbox{p}$, from (0,0) to $(\pi,0)$ (a);
from  $(\pi,0)$ to  $(\pi,\pi)$ (b); and from (0,0) to $(\pi,\pi)$ (c).
The regions in which the line has not been draw have zero imaginary part. }
\label{bimifty.f}
\end{figure}

\begin{figure}
\epsfxsize=8cm
$$
\epsfbox[0 500 596 842]{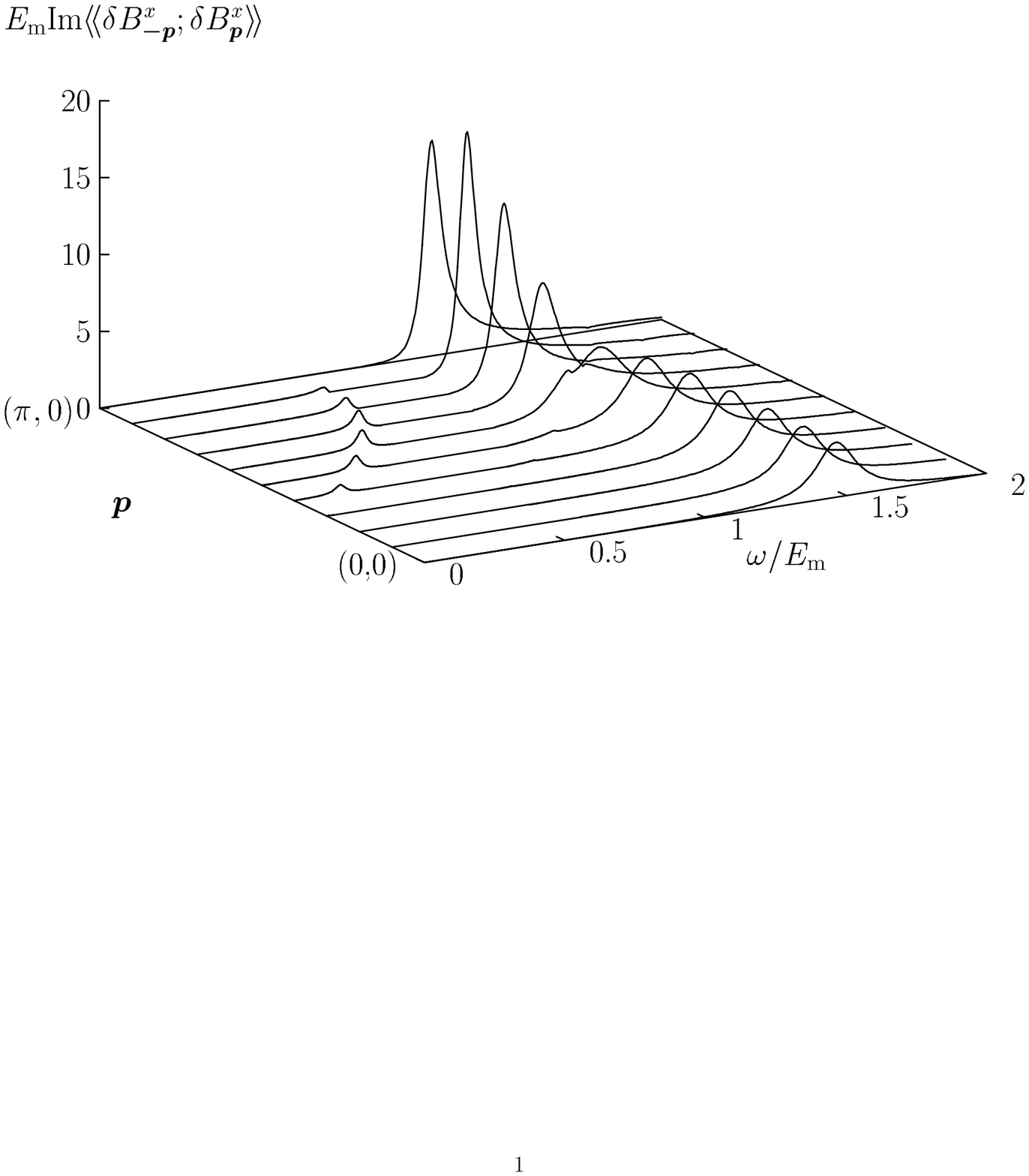}
$$
\epsfxsize=8cm
$$
\epsfbox[0 500 596 842]{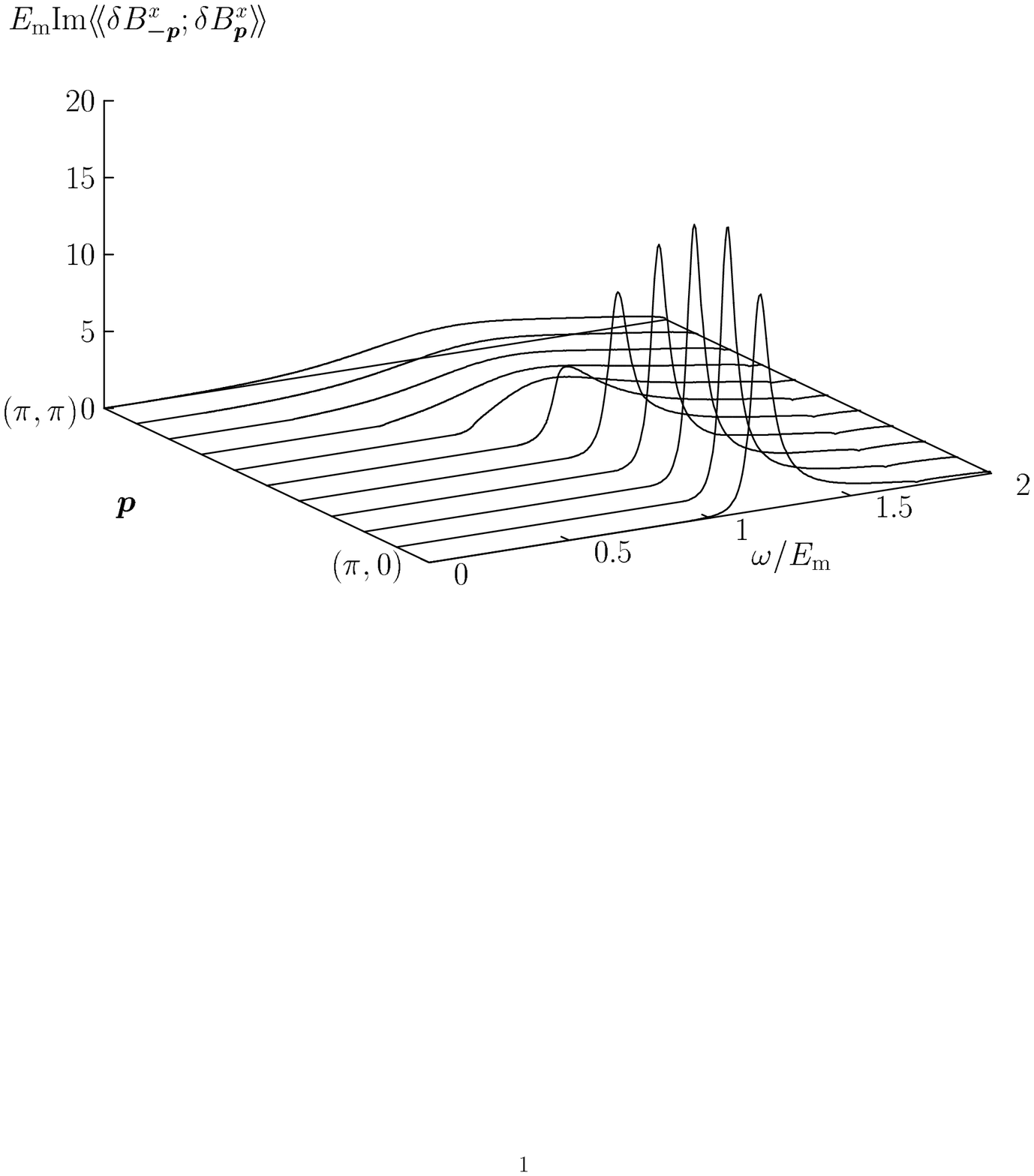}
$$
\epsfxsize=8cm
$$
\epsfbox[0 300 596 842]{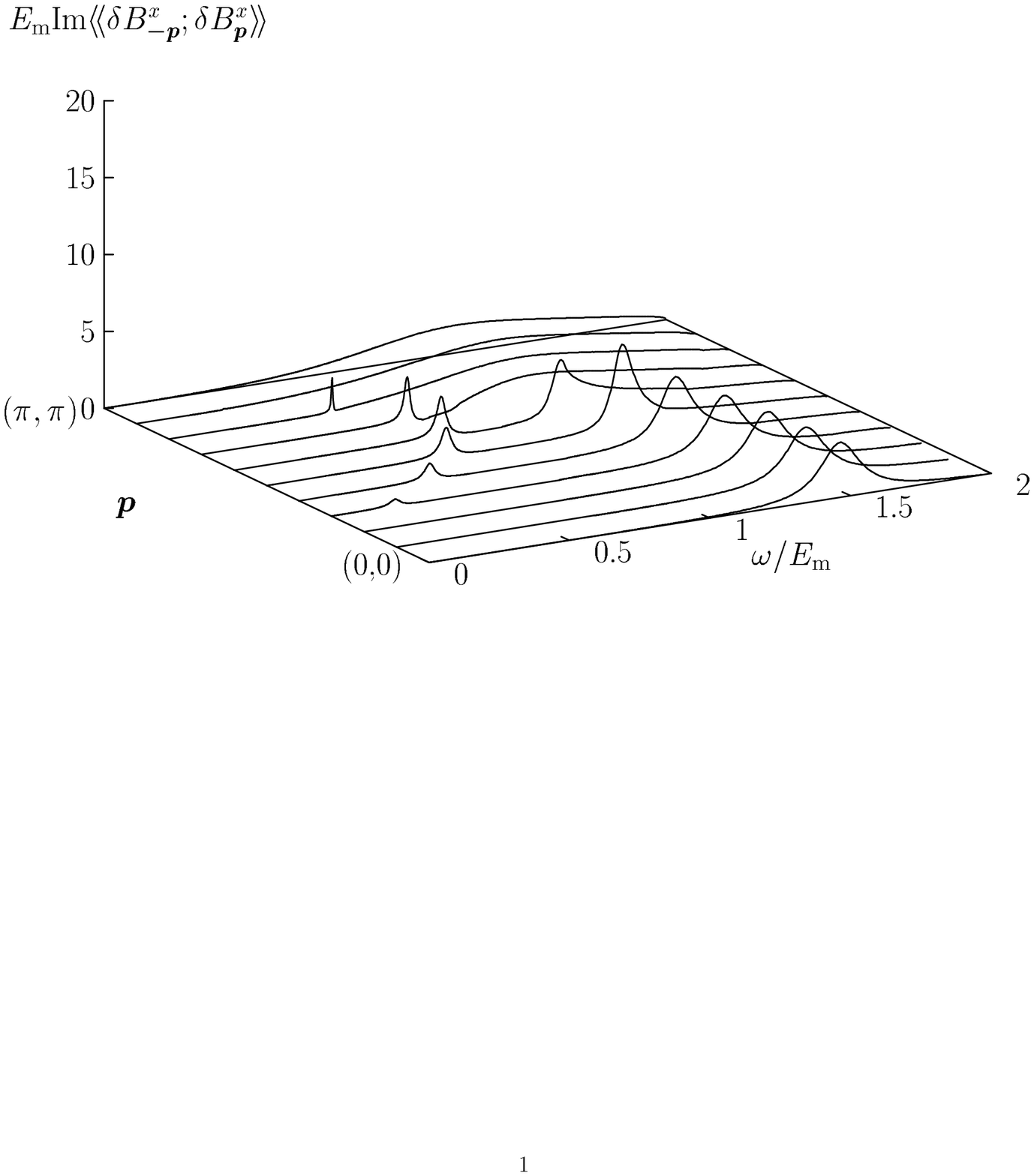}
$$
\caption{Im$\langle\!\langle\delta B^x_{\bbox{ -p}};
\delta B^x_{\bbox{ p}}\rangle\!\rangle$ for $S=1/2$
in the HE approximation in the vertex
as a function of  $\bbox{p}$, from (0,0) to $(\pi,0)$ (a);
from  $(\pi,0)$ to  $(\pi,\pi)$ (b); and from (0,0) to $(\pi,\pi)$ (c).
In the case of real bound states a small imaginary part (0.02 $E_{\rm m}$)
 has been added in the denominator of the Green function. }
\label{bim.f}
\end{figure}

To analyze the following results it is  useful to look
first at the non-interacting ($S=\infty$) case (Fig.~\ref{bimifty.f}).
The regions where the imaginary part is small are good candidates
for having narrow resonances in the interacting case.
 In  Fig.~\ref{bim.f} we show the imaginary part of the Green
function from Eq.~(\ref{bbo})
as a function of the total momentum  $\bbox{p}$ in the high
energy approximation in the vertex but not in the operator
(see Appendix~\ref{hea}).
The bimagnon is only well defined close to $(\pi,0)$.
It disperses upwards on going towards $(0,0)$ [Fig.~\ref{bim.f}(a)]
 and downwards on going towards $(\pi,\pi)$ [Fig.~\ref{bim.f}(b)].
This is shown more clearly in the inset of Fig.~\ref{ade}.
 This indicates that $(\pi,0)$ is a saddle point and hence
it should give a Van Hove singularity when integrated over $\bbox{p}$
[Eq.~(\ref{sdw})]. Because of that the position
 of the peak in the final integrated line shape is the same as for the
$(\pi,0)$ bimagnon.

\begin{figure}
\epsfxsize=8cm
$$
\epsfbox[0 500 596 842]{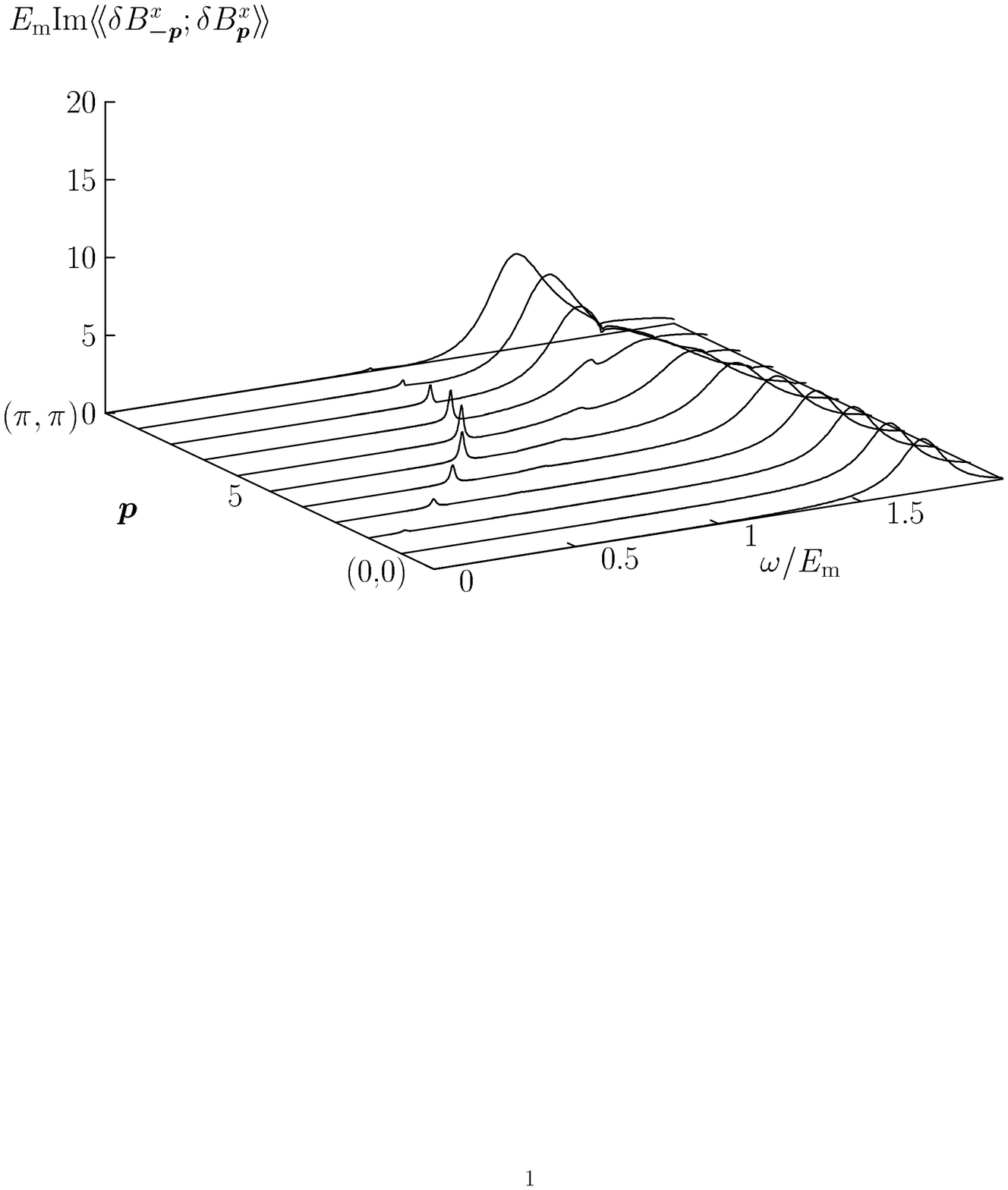}
$$
\epsfxsize=8cm
$$
\epsfbox[0 500 596 842]{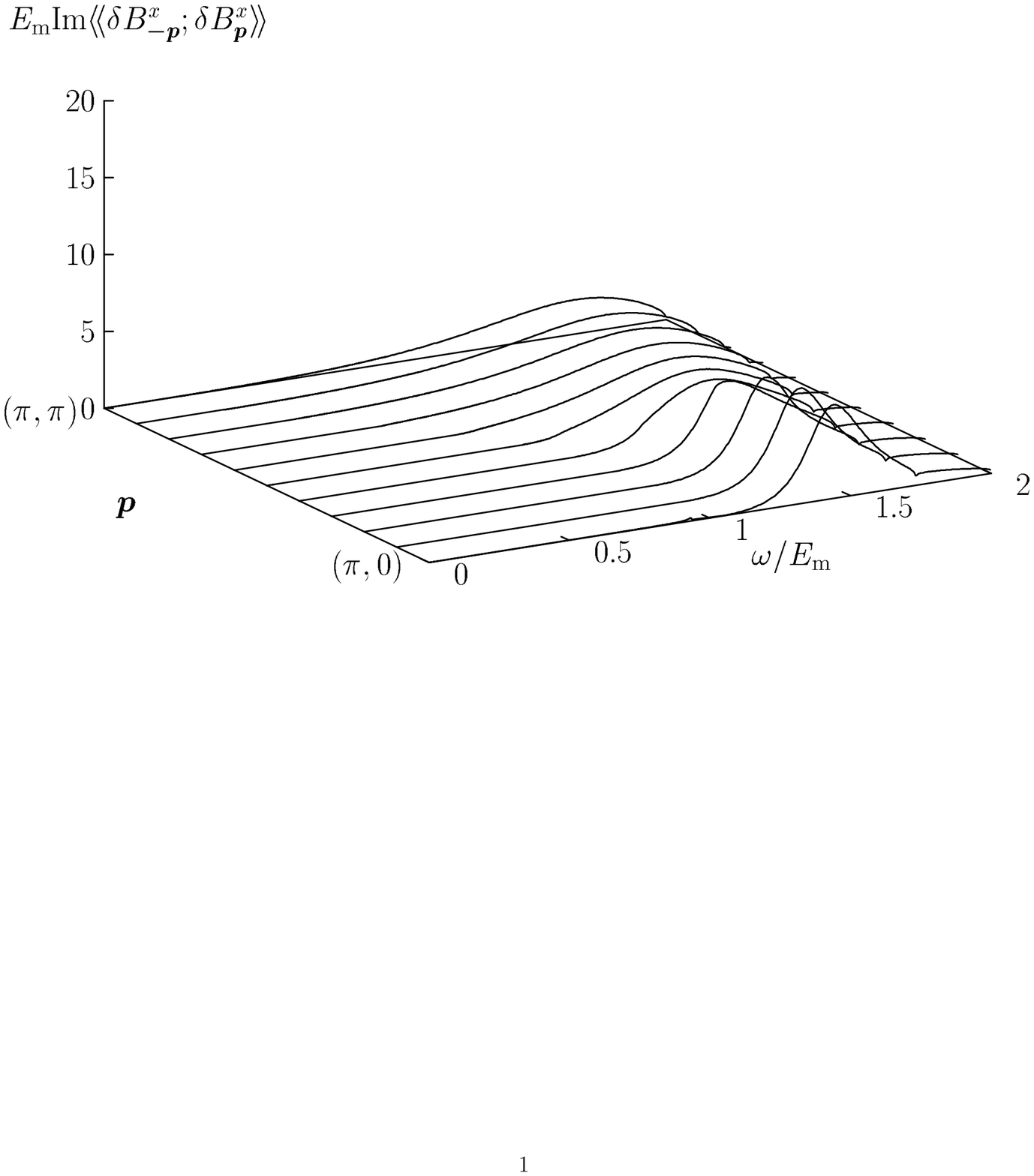}
$$
\epsfxsize=8cm
$$
\epsfbox[0 300 596 842]{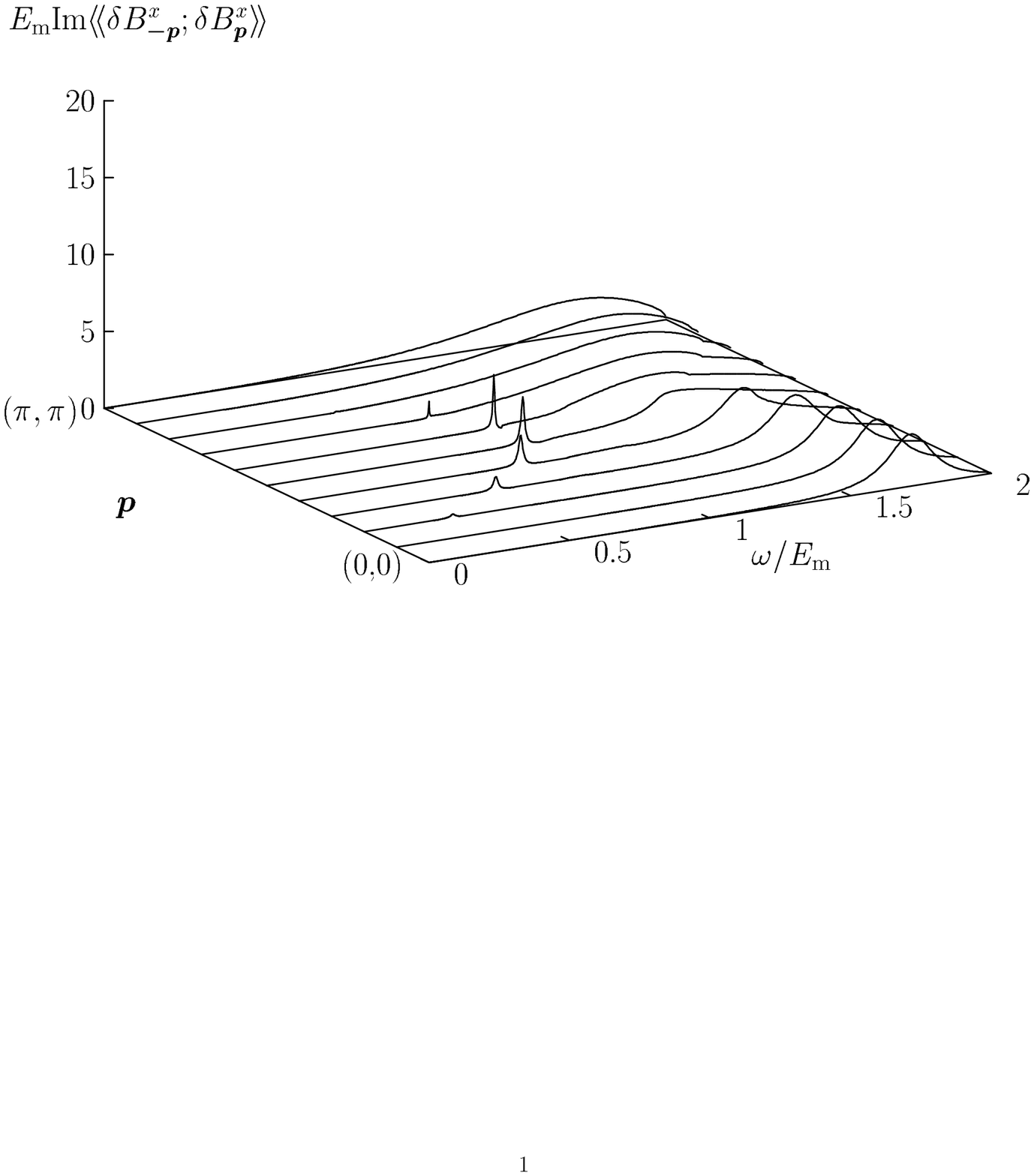}
$$
\caption{Im$\langle\!\langle\delta B^x_{\bbox{ -p}};
\delta B^x_{\bbox{ p}}\rangle\!\rangle$ for $S=1$
in the HE approximation  in the vertex
as a function of  $\bbox{p}$, from (0,0) to $(\pi,0)$ (a);
from  $(\pi,0)$ to  $(\pi,\pi)$ (b); and from (0,0) to $(\pi,\pi)$ (c).
In the case of real bound states a small imaginary part (0.02 $E_{\rm m}$)
 has been added in the denominator of the Green function. }
\label{bim1.f}
\end{figure}

The peak at lower energy is a real bound state. Its position,
 the intensity, and even the existence are
 not very reliable because the position
 is beyond the
range of applicability of the HE approximation.  If the HE
approximation is done in the vertex and in the operator
this peak gets a much larger spectral weight and contributes
spurious intensity to the line shape  (see next Section).
 In  Fig.~\ref{bim1.f}
we show the imaginary part of the Green function
as a function of the total momentum  $\bbox{p}$ for the case of
$S=1$. This will correspond to a material like La$_2$NiO$_4$.
The main bimagnon peak gets overdamped in this case because
the pole shifts up to the region of larger imaginary part
for the noninteracting Green function (Fig.~\ref{bimifty.f}).

\section{COMPARISON WITH EXPERIMENTS}
\label{com}
\subsection{Anisotropy}
\label{ani}
We first discuss the anisotropy because this helps to
fix the values of the effective charges.
For an electric field polarized perpendicular to the plane only
phonons perpendicular to the Cu-O bond contribute. A quick
estimate indicates that the absorption in  this direction is
roughly a factor of 8 smaller than the in-plane contribution.
{}From the experimental side the anisotropy seems to be larger.
One should be aware that the cuprates are in a regime where covalency
is not small with respect to typical gap energies and hence a perturbation
in $t$ is helpful to identify the important processes and discuss
trends, but quantitative estimations are to be taken with care.
The same problem  has been pointed out for the computation of
$J$\cite{esk93}.
 As higher orders in $t$ are included we expect that the anisotropic
contributions grow with respect to the isotropic ones. For example
the charged phonon effects of Fig.~\ref{pro}(b) can become very efficient if
the second hole forms a Zhang-Rice singlet with the hole already present
in Cu$_{\rm L}$ since that process involves a much smaller gap.
Note also that the larger the order in $t$ the longer the range of
the processes that contribute to the anisotropic charges whereas
only local processes contribute to the isotropic charge. Longer
range processes have in general very large form factors.
These effects should give a stronger anisotropy in accordance with the
experiments. This suggest that the process involving
 $\omega_\parallel$ are dominant.  We enforce that in  the
following by taking $q_{\rm I}=0$.

\subsection{IR line shape}

The line shape for phonon-assisted multimagnon absorption is
given by Eq.~(\ref{sdw}). Since the dispersion of the phonons is
very small\cite{rie89} with
respect to the line width we can neglect it and take
Einstein phonons. It is convenient to define the following
function
\begin{eqnarray}
& &I(\omega>0) =- 16\times \nonumber\\
& &\sum_{\bbox{ p}}
\sin^2(\frac{p_x}{2})(\sin^2(\frac{p_y}{2})+ \sin^2(\frac{p_x}{2}))
{\rm Im}\langle\!\langle \delta B^x_{\bbox{ -p}};
\delta B^x_{\bbox{ p}}\rangle\!\rangle
\end{eqnarray}
and $I(\omega<0)=0$. Although the line shape is different for each cuprate
 $I(\omega)$ is the same. The line shape is given by
\begin{equation}
\sigma = \sigma_0 \omega I(\omega-\omega_{\parallel}),
\end{equation}
where
\begin{equation}
\label{sdq}
\sigma_0= \frac{\pi q_{\rm A}^2 }{M V_{\rm Cu} \omega_{\parallel}}
\end{equation}
The absorption coefficient is obtained assuming weak absorption as
\begin{equation}
\alpha=\frac{4\pi}{c\sqrt{\epsilon_1}}\sigma,
\end{equation}
 with $\epsilon_1$ the
real part of the dielectric constant\cite{per94}. We also define
\begin{equation}
\label{si0}
\alpha_0=\frac{4\pi}{c\sqrt{\epsilon_1}}\sigma_0.
\end{equation}

\begin{figure}
\epsfxsize=10cm
$$
\epsfbox[18 144 592 500]{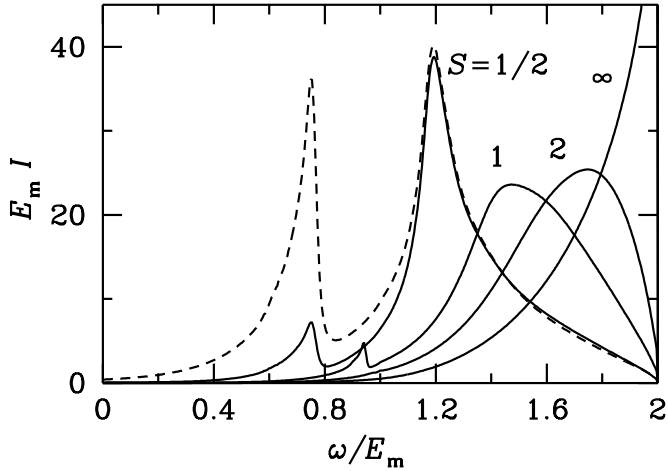}
$$
\caption{$I(\omega)$ for different values of $S$ in the HE approximation
 done in the vertex alone (full line).
 In the case of  $S=1/2$ we show also the result for the HE approximation
 done in the vertex and in the operator (dashed line).
A small imaginary part (0.016$E_{\rm m}$)
 has been added in the denominator of the Green function.}
\label{idw.f}
\end{figure}

In Fig.~\ref{idw.f} we show the function $I(\omega)$ for $S=1/2$
in the HE approximation, done in the vertex alone and in both the  vertex
and  the operator. We see that at HE the two give very similar results
but at lower energy the latter have a much larger feature coming from
the real bound states. Most of this spectral weight is spurious and
can be easily eliminated just by neglecting the contributions from the real
bound states as was done in Ref.~\cite{lor95}. For comparison we show
 $I(\omega)$ for other values of $S$.

To choose $\omega_{\parallel}$ notice that
because of  the Van Hove singularity
the position of the maximum in the line shape is determined
by the position of the bimagnon peak at $(\pi,0)$
 [Eq.~(\ref{emp})] and is given by
 $\omega_{\parallel(\pi,0)}+E^{\rm max}_{(\pi,0)}$.
Because of that, to optimize the position of the maximum, it
is convenient to fix the Einstein phonon frequency
at $\omega_{\parallel}=\omega_{\parallel(\pi,0)}$.

\begin{figure}
\epsfxsize=10cm
$$
\epsfbox[18 0 600 500]{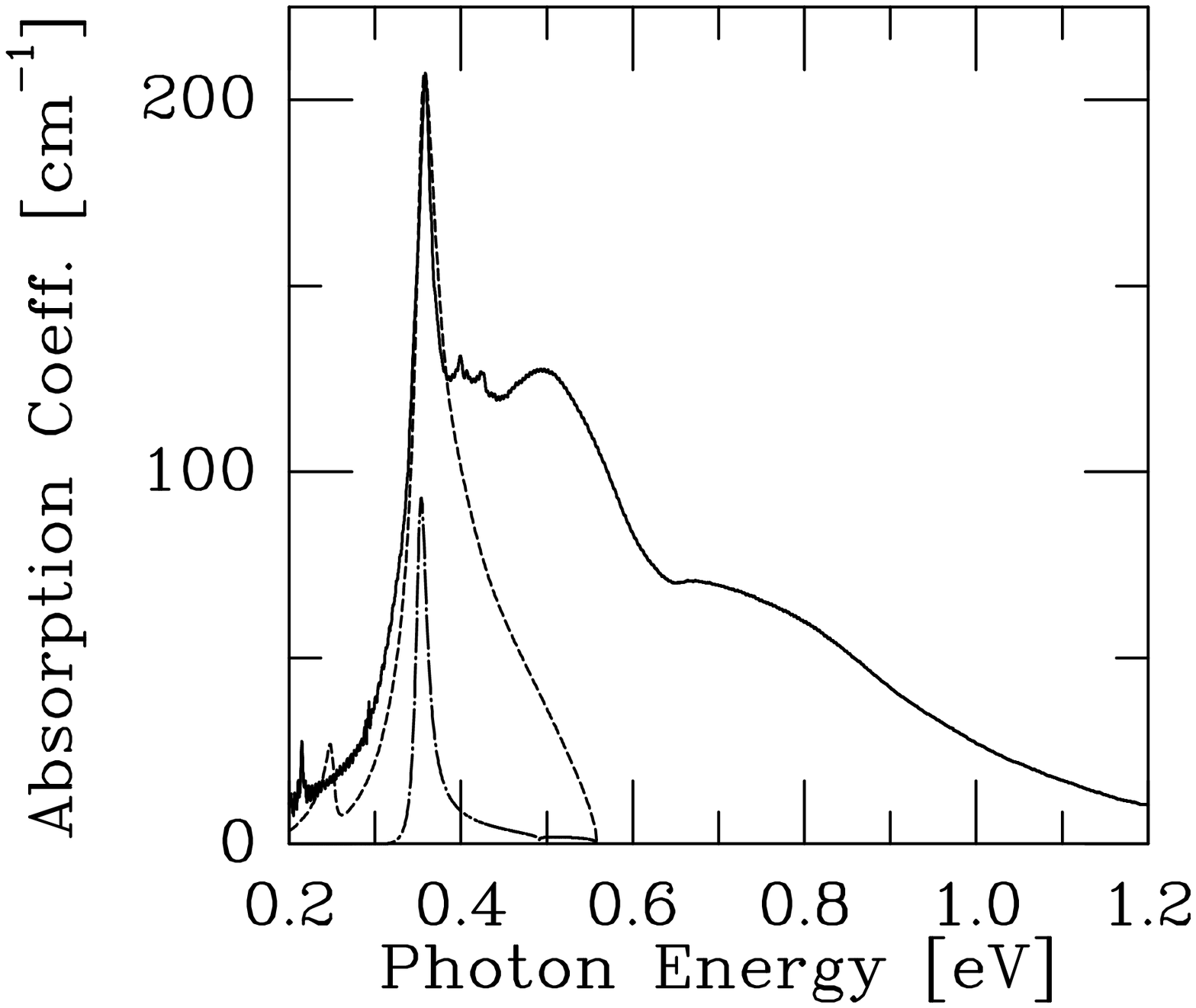}
$$
\caption{Experimental data from Ref.~\protect\cite{per93}
 (solid line) and theoretical
line shape for two-magnon absorption (dashed line) in Sr$_2$CuO$_2$Cl$_2$.
 The dashed doted line is the contribution to the line shape from the
bimagnon at  $\bbox{p}=(\pi,0)$. Parameters are
$\omega_\parallel=0.061$eV $J=0.107$eV, $\alpha_0=3.72$cm$^{-1}$ }
\label{asr}
\end{figure}

The dashed curve in  Fig.~\ref{asr} shows the theoretical
and experimental line shape in the HE approximation (vertex only)
for Sr$_2$CuO$_2$Cl$_2$.
Here $\omega_{\parallel(\pi,0)}$ has not been measured so we
took it as 5\% smaller than the measured IR frequency\cite{taj91}
of the E$_{u}$ stretching mode as
suggesting by comparing with La$_2$CuO$_4$.
  Notice that these differences
are insignificant anyway. The fitting for the primary peak is quite good.
The improved theory (HE approximation in the vertex alone)
 has a slightly narrower line shape which makes the fit not as good as
for the more approximate line shape of Ref.~\cite{lor95}.
Still the fit is quite good and this is surprising because
such a good fit, especially for the width was not possible within RPA
in the Raman case\cite{sin89}. In fact the experimental Raman line
shape is much broader than the theoretical prediction shown in
Fig.~\ref{ram.f}. This suggest that RPA is not so accurate in the
case of $S=1/2$.  The fact  that we obtain a reasonable
fit in our case can be partially reconciled with the relatively
bad performance of RPA in the Raman case
by the fact that a structure that is artificially
broadened around $p=(\pi,0)$ by a factor of 2
 (but still much narrower than the integrated line shape) does not
 change significantly the final result.
i.e. the final width in the line shape is not so sensitive
to the width of the bimagnon close to $p=(\pi,0)$ so this experiment
does not prove or disprove the accuracy of RPA for a particular momentum.

{}From the position of the maximum we  found
$J=0.107eV$. An
alternative way to estimate $J$ is from the position
of the maximum in the Raman peak, Eq.~(\ref{em0}). From the measurements
of Ref.~\cite{tok90} we found $J=0.103eV$.
  In this estimation one should again  be aware that the accuracy of
RPA for a $S=1/2$ system has been questioned\cite{sin89}.

\begin{figure}
\epsfxsize=10cm
$$
\epsfbox[18 0 600 500]{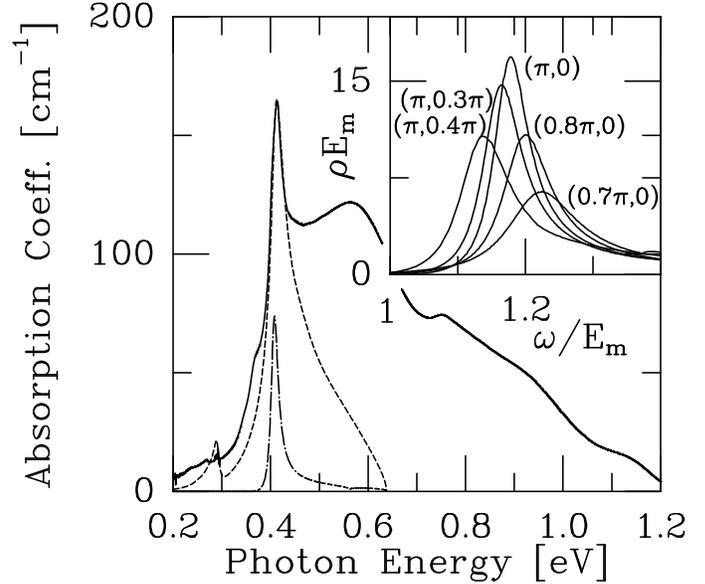}
$$
\caption{Experimental data from Ref.~\protect\cite{per93} (solid line)
and theoretical line shape for two-magnon absorption (dashed line) in
La$_2$CuO$_4$.
 The dashed doted line is the contribution to the line shape from the
bimagnon at  $\bbox{p}=(\pi,0)$
 Parameters are
$\omega_\parallel=0.080$eV\protect\cite{rie89},
 $J=0.121$eV, $\alpha_0=2.88$cm$^{-1}$ }
\label{ade}
\end{figure}

 We keep for comparison  in Fig's.~\ref{asr},\ref{ade}
 the contribution
from the real bound states although their position,  intensity
and even their existence
are not very reliable because they are beyond the applicability
of the HE approximation.

The data shows also tiny structures above the primary
peck. These structures can also be  due to adsorbates\cite{percom}
or, as the same authors suggest, they can be phonon side bands.
In fact the two-phonon plus bimagnon contribution is expected at
$E^{\rm max}_{(\pi,0)}$ plus two phonon frequencies. Its spectral weight is
 much smaller than the one-phonon plus bimagnon.
 For example
the isotropic charge is of order $\xi_{\rm I}\sim
\frac{\beta}{\Delta}
 q_{\rm I}$ and  hence the absorption is of order
$\langle\!\langle P_{\rm 2ph+mag}; P_{\rm 2ph+mag} \rangle\!\rangle \sim
\frac{\langle u_{\bbox{i}+\bbox{\delta}/2}^2\rangle}{a_{pd}^2}
\langle\!\langle P_{\rm 1ph+mag}; P_{\rm 1ph+mag} \rangle\!\rangle$, where
we have taken a factor $\frac{\beta^2 a_{pd}^2}{\Delta^2}\sim 1$. This is
roughly three orders of magnitude smaller than the previous contributions
and can explain the tiny structures reported for Sr$_2$CuO$_2$Cl$_2$.
 The shift from the
primary peak is $\sim$ 0.04 eV and 0.07eV in good agreement with typical
values of  $\omega_\perp$ and $\omega_\parallel$.
 Note that only the
presence of a sharp bimagnon state would make those processes observable.

In Fig.~\ref{ade} we show the fit for La$_2$CuO$_4$. Here we
found $J_0=0.121$eV. The corresponding value from the position
of the maximum in the observed Raman line shape\cite{tok90,sin89}
is $J=0.118$eV. The same comment
as before applies for the accuracy of this RPA estimates. The
superexchange was estimated also from a study of the moments
of the Raman line shape\cite{sin89} with the result $J=0.128$eV.

 The $\omega_\parallel$ phonon is very anomalous in
orthorombic La$_2$CuO$_4$ since it splits due to anharmonicities\cite{rie89},
its partner being at $\omega'_\parallel=$0.06eV at room temperature.
Presumably this produces the shoulder observed at lower energies in
the experiments (Fig.~\ref{ade}), although the distance to
the primary is larger than expected which may be because other phonons are
 involved.   This feature was assigned to direct two-magnon
absorption \cite{per93,per94} made weakly allowed by the lower lattice
symmetry according to the results of Ref.~\cite{tan65}.
However we found that the dipole moment for this process is directed
in the direction bisecting an angle made by the Cu-O-Cu bond\cite{mor66}
and hence can only contribute for a field perpendicular to the plane.
 It is possible that a soft mode of the distorted structure gives rise
to this effect since this will generate an enhanced intensity at lower
frequencies according to Eq.~(\ref{sdw}).
although more theoretical and experimental work
is needed to clarify this point. Note that the shift
between the primary peak and the shoulder is in the range of
phonon energies ($\sim 0.04$ eV).

\subsection{Spectral weights}

The measured spectral weight for the primary peck is determined by the
parameter $\alpha_0$. With Eq.~(\ref{si0}) we can converted that
to $\sigma_0$. For Sr$_2$CuO$_2$Cl$_2$ we have $\epsilon=5.5$\cite{taj91}
and we get $\sigma_0=0.023 \Omega^{-1}$cm$^{-1}$ and for La$_2$CuO$_4$
$\epsilon=6.0$\cite{taj91} and $\sigma_0=0.018 \Omega^{-1}$cm$^{-1}$.
We can use Eq.~(\ref{sdq}) to estimate $q_{\rm A}$. For La$_2$CuO$_4$,
$V_{\rm Cu}=95\times 10^{-24} $cm$^3$ and we approximate $M$ with an O mass
(a reduced  mass would be more appropriate). We get $q_{\rm A}/e=0.082$.
For Sr$_2$CuO$_2$Cl$_2$ with  $V_{\rm Cu}=123\times 10^{-24} $cm$^3$ we get
$q_{\rm A}/e=0.088$. Examining the expression for the effective charge
we find that is of order
$q_{\rm A}/e\sim\frac{2JU_{pd}}{\Delta^2} \sim 0.1$ in very good agreement
with the experimental values. One should take into account however that
the previous estimates neglect the weight in the side bands so that
the observed spectral weight is in reality larger than the one
estimated with $q_{\rm A}/e\sim 0.1$. This leaves room for the effects
discussed in Sec.~\ref{ani}.





\subsection{Prediction for La$_2$NiO$_4$}

\begin{figure}
\epsfxsize=10cm
$$
\epsfbox[18 144 592 600]{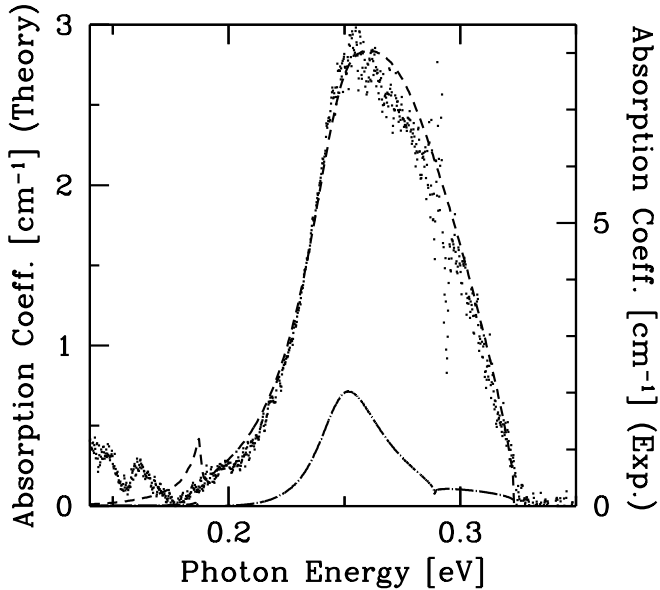}
$$
\caption{Theoretical line shape for two-magnon absorption in
La$_2$NiO$_4$ (dashed line).
 The dashed doted line is the contribution to the line shape from the
bimagnon at  $\bbox{p}=(\pi,0)$
 Parameters are $\omega_\parallel=0.066$eV\protect\cite{pin89}, $J=0.030$eV,
 $\alpha_0=6.0\times 10^{-2}$cm$^{-1}$. The dots are unpublish experimental
data\protect\cite{per95}. }
\label{nio}
\end{figure}

In this case structural parameters are similar  to the previous
compound, $V_{\rm Ni}=94\times 10^{-24} $cm$^3$ and $\epsilon_1=5.4$
however the ratio $J/\Delta$ is an order of magnitude smaller\cite{sug90}.
So we expect $q_{\rm A}\sim\frac{2JU_{pd}}{\Delta^2} \sim 0.01$
 which makes the absorption two orders of magnitude
smaller than in the previous compounds. With this value of
$q_{\rm A}$ we get  $\sigma_0=3.7\times 10^{-4} \Omega^{-1}$cm$^{-1}$.
We have predicted
the line shape taking the phonon frequency from neutron
scattering experiments ($\omega_\parallel=0.066$eV\cite{pin89})
and the superexchange form Raman experiments ($J=0.030$eV\cite{sug90}).
The expect absorption is plotted in Fig.~\ref{nio}. As explained
before there are no sharp features in the $S=1$ case.

After this work has been completed we send the  predicted line shape
to J. Perkins and in response
we received from them the experimental data plotted in Fig.~\ref{nio}.
The agreement is remarkably good for the line shape and our rough
 estimation of the oscillator  strength happened to be of the correct
order of magnitude. Comparison with the previous case  illustrates
quite obviously the remarkably difference between the $S=1$ and the
$S=1/2$ system the latter showing the strong side bands which
require an explanation beyond RPA\cite{lor95a}.

\section{DISCUSSION AND CONCLUSIONS}
\label{conc}
The experiments\cite{fal93}
 show also very strong side bands. These are expected
 as the effect of
quantum fluctuations corresponding to the creation of an arbitrary
even number of magnons. This is in principle also true for the Raman
case and in fact a structure
attributed to a four magnon  process was observed in Raman\cite{sug90}
at a similar energy to the first side band
(note that phonon energies are almost negligible here).

In the Ising limit one can roughly estimate the position of these side bands.
The minimum energy to excite 4 magnons
(plaquette configuration) is $\omega_{\rm 4mag}=4J$ which correspond to
the first side band (see Fig's.~\ref{asr},\ref{ade}). The next
side band is attributed to a process in which four spins are flipped in a
column which has an excitation energy of $\omega_{\rm 4mag}=5J$ in the Ising
limit. In general one can show that side bands are expected at integer values
of $J$ as observed experimentally.
 We have also computed the side bands by
exact diagonalization of a small cluster\cite{lor95a}.
 The exact result confirms
the Green function calculation and show the side
bands corresponding to higher multimagnon process.
The relative weight of the side bands seem to be smaller
than in the experiments presumably because of
finite site effects or the presence of other process
in the magnetic Hamiltonian as in Raman\cite{rog89,hon91}.
Exact diagonalization
calculations of the Raman spectra\cite{gag90} show also
this kind of side bands.

As the temperature is raised one expects that the structure
broadens in a way similar to the Raman line shape and at the same time
a hot phonon band should appear at $E^{\rm max}_{(\pi,0)}-\omega_{\parallel}$
corresponding to a process in which a phonon of the bath
 is absorbed and a bimagnon is emitted. It would be interesting if
 this effect can be seen experimentally.

It is interesting to point out that one can in principle see the bimagnons
for a particular momentum without the assistance of the phonon
in neutron scattering experiments\cite{cow69} or high resolution
electron energy loss (HREELS) experiments.
 In the latter for energy losses small with respect to the energy of the
 incident electrons  and in specular scattering geometry,
 dipole selection rules are valid
and one should also see the analogue of the phonon-assisted
IR line shape.
Preliminary HREELS results\cite{pot95} in Sr$_2$CuO$_2$Cl$_2$ show a structure
which agrees very well with Perkins {\it et al.} data.
Other possibility to see this excitations without the assistance of the phonon
would be to add impurities. In this way the zero total momentum
selection rule will
not apply any more and the narrow $(\pi,0)$ bimagnon should give some signature
at $E^{\rm max}_{(\pi,0)}$. The impurity should also break inversion
symmetry (which to same extent is always does)
in order to make the two magnon excitations IR active
and they should not introduce doping otherwise the doping dependent mid-IR
band\cite{uch91} would mask everything. Possible candidates would be
   Nd$_{2-x}$ Pr$_{x}$ CuO$_4$ or similar combinations with
other rare earths. Since the Nd is on top of a Cu
the O is not any more in a center of inversion.

Another interesting experiment would be to repeat Perkins {\it et al.}
 experiment in
 an insulator with in-plane anisotropy. Since an electric field in the
 $x$ direction couples with bimagnons at  $(\pi,0)$  and one in the $y$
direction with bimagnons at  $(0,\pi)$,
 if this two directions are not equivalent the
primary peak should split in a twined crystal.


We do not believe that  this mid-IR excitations are related to the
doping dependent mid-IR bands observed in the cuprates for which a completely
different mechanism has been proposed\cite{lor93a,lor94a}. For example
 for Nd$_2$CuO$_{4-y}$ in the notation of Ref.~\cite{tho91} the band
labeled
$J$ at 0.16eV  has been quite undoubtedly  shown to be due to polaron
formation\cite{yon92,yon93,dob94,lor94} by Calvani {\it et al.}\cite{cal94}.
The structure labeled $I$ at 0.76eV
is the analogue of the mid-IR band reported by Uchida {\it et al.}\cite{uch91}
for La$_{2-x}$Sr$_x$CuO$_4$ at 0.5eV and can be explained
by a purely electronic mechanism\cite{lor93a,lor94a}. Only the much weaker
structure labeled $K$ at 0.34eV may be related to this excitations.

In this work we have computed effective coupling constants of light with
multimagnon excitations assisted by phonons and the line shape of the primary
peak.  Our results explain recent
measured absorption bands in the mid IR of parent cuprate superconductors
and show that this technique proves high energy magnetic excitations.
This is very interesting because IR spectroscopy is a
 technique  intrinsically  more accurate and with a much better signal to
noise ratio than other techniques available like Raman or neutron scattering.
We have  demonstrated the existence of very sharp virtual bound states
of  magnons in spin 1/2 systems at momentum $(\pi,0)$ and show
the shape of the excitations at different momentum. To the best of our
 knowledge is the first time that the RPA equations for the two magnon problem
are presented for arbitrary total momentum.

\acknowledgments

We acknowledge the authors of Ref.~\cite{per93} for sending us their
work previous to publication and for enlightening discussions and
R. Eder and  M. Meinders for helping us with the exact
diagonalization calculations.
This investigation was supported by the Netherlands Foundation for
Fundamental Research on Matter (FOM) with financial support from the
 Netherlands Organization for the Advance of Pure Research (NWO)
and Stichting Nationale Computer Faciliteiten (NCF).
Computations where perform at SARA (Amsterdam).
 J.L. is supported  by  a postdoctoral fellowship
granted by the  Commission of the European Communities.

\appendix
\section{Vertex functions}
\label{ver.a}
The symmetrized vertex in Eq.~(\ref{vres})  are given by
$\Gamma= \Gamma^{\parallel} + \Gamma^{\perp}$ with
\begin{eqnarray}
&\Gamma&_{1234}^{\parallel}=-\frac{Jz}4
[  \gamma_{2+1} (u_1v_2v_3u_4+v_1u_2u_3v_4)\nonumber\\
&+&\gamma_{2-4} (u_1u_2u_3u_4+v_1v_2v_3v_4+u_1v_2u_3v_4+v_1u_2v_3u_4)
\nonumber\\
 &+& 1\leftrightarrow 2 + 3\leftrightarrow 4 +
 1\leftrightarrow 2 \  {\rm and}\  3\leftrightarrow 4],\\
&\Gamma&_{1234}^{\perp}=-\frac{Jz}8[(2\gamma_{2}+\gamma_{3})u_1u_2v_3u_4
 +  (2\gamma_{1}+\gamma_{3}) v_1v_2u_3v_4 \nonumber \\
&+& (2\gamma_{3}+\gamma_{2}) u_1v_2u_3u_4
 +  (2\gamma_{4}+\gamma_{2})v_1u_2v_3v_4 \nonumber\\
 &+& 1\leftrightarrow 2 + 3\leftrightarrow 4 +
 1\leftrightarrow 2 \  {\rm and}\  3\leftrightarrow 4 ].
\end{eqnarray}
In term of total momentum  and  relative momentum variables
the vertex in Eq.~(\ref{gsw}) can be rewritten as
\FL
\begin{eqnarray}
\label{ver}
\Gamma_{\bbox{p}\bbox{q}_1\bbox{q}_2}^{\parallel}&=&-J[\sum_{\delta l=1,2}
(f^{l\delta}_{\bbox{p}\bbox{q}_1}f^{l\delta}_{\bbox{p}\bbox{q}_2}
+h^{l\delta}_{\bbox{p}\bbox{q}_1}h^{l\delta}_{\bbox{p}\bbox{q}_2})+
\frac{z}2\gamma_{\bbox{p}}f^3_{\bbox{p}\bbox{q}_1}f^3_{\bbox{p}\bbox{q}_2}],
\nonumber\\
& &\\
\Gamma_{\bbox{p}\bbox{q}_1\bbox{q}_2}^{\perp}&=&-J\sum_{\delta}
[ \cos(\frac{p_{\delta}}2)(
f^{1\delta}_{\bbox{p}\bbox{q}_1}f^3_{\bbox{p}\bbox{q}_2}+
\frac12 h^{1\delta}_{\bbox{p}\bbox{q}_1} h^3_{\bbox{p}\bbox{q}_2})\nonumber\\
&+&\frac12 \sin(\frac{p_{\delta}}2) h^{2\delta}_{\bbox{p}\bbox{q}_1}
h^4_{\bbox{p}\bbox{q}_2}  + \bbox{q}_1 \leftrightarrow \bbox{q}_2 ]\nonumber
\end{eqnarray}
where we used the definitions in Eqs.~(\ref{fld}),(\ref{hld}).

\section{Solution of the two magnon problem for 2d and \lowercase{$p_x=p_y$}}
\label{pxpy}

In this case the Green functions for fixed $\bbox{ p}$
are invariant with respect to the exchange
$q_x \leftrightarrow q_y$. As is done in the theory
of Raman scattering we can define
$G_{l,l'}^d=2(G_{lx,l'x}-G_{lx,l'y})$
and a similar definition for the noninteracting
Green functions.
In this case Eq.~(\ref{gdi}) reduce to a $2\times 2$ problem,
\begin{equation}
G_{ll'}^d =G_{ll'}^{d(0)}-\frac{J}2
\sum_{l''=1,2}G^{d(0)}_{ll''} G_{l''l'}^d
\end{equation}
which can be solved for
\FL
\begin{eqnarray}
\label{gd}
& &G_{11}^d= \\
& &\frac{G^{d(0)}_{11}
+\frac{J}2 (G^{d(0)}_{11}G^{d(0)}_{22}-(G^{d(0)}_{12})^2)}
{1+\frac{J}2 (G^{d(0)}_{11}+G^{d(0)}_{22})
+\frac{J^2}4(G^{d(0)}_{11}G^{d(0)}_{22}-(G^{d(0)}_{12})^2)}\nonumber.
\end{eqnarray}

\section{Solution of the two magnon problem for
 \lowercase{$\bbox{p}=(\pi,0,0..)$} }
\label{px0}
In this case the Green function for the $x$ direction
decouples from the rest in Eq.~(\ref{gdi}).
This is because $G^{(0)}_{lx,3}=0$,
 $ G^{(0)}_{lx,l'\delta}=0$ with $\delta\ne x$.
 To see this notice that in the
definition Eq.~(\ref{gi0}), the kernel changes sign under the replacement
$\bbox{q}\rightarrow \bbox{q}+(\pi,0,0..) $. So we get,
\begin{equation}
G_{lxl'x} =G_{lxl'x}^{(0)}-J
\sum_{l''=1,2}G^{(0)}_{lxl''x} G_{l''xl'x}
\end{equation}
we can solve for
\FL
\begin{eqnarray}
\label{gxx}
& & G_{1x1x(\pi,0,0...)}= \\
& &\frac{G^{(0)}_{1x1x}+G^{(0)}_{2x2x}+2G^{(0)}_{1x2x}
+4J(G^{(0)}_{1x1x}G^{(0)}_{2x2x}-(G^{(0)}_{1x2x})^2)}
{1+2J(G^{(0)}_{1x1x}+G^{(0)}_{2x2x})
+4J^2(G^{(0)}_{1x1x}G^{(0)}_{2x2x}-(G^{(0)}_{1x2x})^2)}\nonumber.
\end{eqnarray}

\section{High energy approximation}
\label{hea}
When  the line shape is
positioned at  sufficiently high energies ($\omega >E_{\rm m}$)
it is a good approximation to neglect the $v$'s respect to the $u$'s
in Eqs.~(\ref{alp}-\ref{hld}). In principle this can be done
everywhere in the expressions for the Green functions.
However it is convenient to do this in two steps.
First we define
\begin{equation}
f^{\pm\delta}_{\bbox{p}\bbox{q}}=f^{1\delta}_{\bbox{p}\bbox{q}}\pm
f^{2\delta}_{\bbox{p}\bbox{q}}
\end{equation}
and we do the HE approximation in the vertex, i.e. we neglect
all contributions that involve  $v$'s in  Eq.~(\ref{ver}).
We denote with  primes the quantities computed in the HE approximation, i.e.
\begin{eqnarray}
\Gamma_{\bbox{p}\bbox{q}_1\bbox{q}_2}^{'\parallel}&=&-2J\sum_{\delta}
f^{+\delta}_{\bbox{p}\bbox{q}_1}f^{+\delta}_{\bbox{p}\bbox{q}_2}
\nonumber\\
& &\\
\Gamma_{\bbox{p}\bbox{q}_1\bbox{q}_2}^{'\perp}&=&0\nonumber
\end{eqnarray}
Now we rederive  Eq.~(\ref{gdi}) in the HE approximation.
\begin{equation}
G'_{\mu \nu} =G_{\mu\nu }^{(0)}-2J
\sum_{\delta}G^{(0)}_{\mu,+\delta} G'_{+\delta,\nu}.
\end{equation}
We can treat $\delta$ as a matrix or vector index
of dimension d and put this as
\begin{equation}
\label{gll}
G'_{ll'} =G_{ll'}^{(0)}-2JG^{(0)}_{l+} G'_{+l'}.
\end{equation}
Notice that $G'_{11}$, $G'_{22}$, $G'_{\pm\pm}$ are matrixes
$G'_{13}$, $G'_{31}$ are vectors and $G'_{33}$ is a scalar.
Solving the  d$\times$d problem for $G'_{+l}$ we get
\begin{equation}
\label{g+l}
G'_{+l} =(1+2JG^{(0)}_{++})^{-1} G_{+l}^{(0)}.
\end{equation}
and replacing in Eq.~(\ref{gll}) we get also $G'_{ll'}$.
We also have the exact transformation
\begin{mathletters}
\label{tra}
\begin{eqnarray}
G'_{11}&=&\frac14 (G'_{++}+G'_{--}+2G'_{+-}),\label{gp11}\\
G'_{31}&=&\frac12 (G'_{3+}+G'_{3-}),\\
G'_{13}&=&\frac12 (G'_{+3}+G'_{-3}).
\end{eqnarray}
\end{mathletters}
When calculating the Green function for the two magnon operator,
Eq.~(\ref{bbo}) one would be tempted to make the same approximation and
keep only $G'_{++}$ since according to the definitions Eqs.~(\ref{albe}),
(\ref{fld}),(\ref{gmn}), all other contributions are negligible.
This means that not only the vertex
or equivalently the interaction part of the Hamiltonian
 ($V_{\rm res}^{\rm RPA}$) is approximated by ($V_{\rm res}^{' \rm RPA}$)
but also the operator $\delta B^x_{\bbox{p}}$ is approximated
by  $\delta B'^x_{\bbox{p}}$. We call this the HE approximation
in the vertex {\em and} in the operator.  Although this is perfectly
consistent at high energies this approach produces a large
spurious contribution at low energies.  To see this
consider the case of $\bbox{p}=0$. We can define
\begin{eqnarray}
&\delta B^s_0=& \delta B^x_0+ \delta B^y_0,\nonumber\\
& &\\
&\delta B^d_0=& \delta B^x_0- \delta B^y_0\nonumber\\
\end{eqnarray}
Now we find,
\begin{eqnarray}
\langle\!\langle \delta B^x_0;\delta B^x_0\rangle\!\rangle=
\frac14 &(&\langle\!\langle \delta B^d_0;\delta B^d_0\rangle\!\rangle
+\langle\!\langle \delta B^s_0;\delta B^d_0\rangle\!\rangle\nonumber\\
&+&\langle\!\langle \delta B^d_0;\delta B^s_0\rangle\!\rangle
+\langle\!\langle \delta B^s_0;\delta B^s_0\rangle\!\rangle).
\end{eqnarray}
Only the first term contributes to the line shape.
i.e the $\bbox{p}=0$ contribution to the IR line shape
is identical to the Raman line shape.   This follows from the fact
that $B^s_0$ is proportional to the Hamiltonian and hence it commutes with it.
To enforce that in spin-wave theory, we would like  $B^s_0$ to commute
with $H$ order by order in $1/S$. However if we do the high energy
approximation in the operator  of Eq.~(\ref{bbo}), we miss some terms of
order $S$ in $B^s_0$ and then we find that to order
$S^2$, $[\delta B'^s_{\bbox{p}},H_0]\ne 0$. This gives spurious
scattering in the $s$ channel.
 We can avoid that by doing
the high energy approximation in the vertex alone. In this case
$G'_{13}$, $G'_{31}$ and $G'_{33}$ are kept in Eq.~(\ref{bbo}) and
computed with Eqs.~(\ref{gll}),(\ref{g+l}) and (\ref{tra}).
 It is instructive to check
analytically that in this way the non-interacting
 $\langle\!\langle \delta B^s_0;\delta B^s_0\rangle\!\rangle^0=0$.

In the 2-d case we find,
\begin{equation}
 (1+2JG^{(0)}_{++})^{-1} =\frac1\Delta_{++}
\left(\begin{array}{cc}
 1+2J G^{(0)}_{y+y+} & -2J G^{(0)}_{x+y+}\\
 -2J G^{(0)}_{x+y+}  & 1+2J G^{(0)}_{x+x+}\end{array} \right)
\end{equation}
with
\begin{eqnarray}
\Delta_{++}= 1& +& 2J (G^{(0)}_{x+x+}+G^{(0)}_{y+y+}) \nonumber\\
& +& 4J^2 [G^{(0)}_{x+x+}G^{(0)}_{y+y+}-(G^{(0)}_{x+y+})^2 ].
\end{eqnarray}
so we get for $G'_{+x+x}$
\FL
\begin{equation}
\label{bb}
G'_{+x+x}=
\frac{ G^{(0)}_{+x+x}+2J[G^{(0)}_{+x+x}G^{(0)}_{+y+y}-(G^{(0)}_{+x+y})^2] }
{\Delta_{++}}
\end{equation}
 This reduce to Eq.~(\ref{gpi0}) for $\bbox{p}=(\pi,0)$ using the properties
of Appendix~\ref{px0}.


\end{document}